\documentclass[twocolumn,aps,prd,floatfix,nofootinbib,superscriptaddress]{revtex4-2}
\usepackage[utf8]{inputenc}
\usepackage{graphicx}
\usepackage{graphics}
\usepackage{amssymb,amsmath}
\usepackage{tabularx}

\begin{document}        

\title{\boldmath Measurement of the $e^+e^-\to K_SK_L$ cross section
near the $\phi(1020)$ resonance with the SND detector}

\begin{abstract}
The cross section for the $e^+e^-\to K_SK_L$ process is measured in
the center-of-mass energy range from 1000 MeV to 1100 MeV in the experiment
with the SND detector at the VEPP-2000 $e^+e^-$ collider. The measurement
is carried out in the $K_S\to 2\pi^0$ decay mode. Data with an integrated 
luminosity of 20 pb$^{-1}$ recorded in 2018 at 18 energy points are used
in the analysis. The systematic uncertainty in the measured cross section 
at the maximum of the $\phi$ resonance is 0.9\%. The mass, width of the
$\phi$ meson, and the product of the branching fractions
$B(\phi\to K_SK_L)B(\phi\to e^+e^-)$ are determined from
the fit to the cross-section energy dependence.
\end{abstract}

\author{M.~N.~Achasov}
\affiliation{Budker Institute of Nuclear Physics, SB RAS, Novosibirsk, 630090, Russia}
\affiliation{Novosibirsk State University, Novosibirsk, 630090, Russia}
\author{A.~Yu.~Barnyakov}
\affiliation{Budker Institute of Nuclear Physics, SB RAS, Novosibirsk, 630090, Russia}
\affiliation{Novosibirsk State University, Novosibirsk, 630090, Russia}
\author{E.~V.~Bedarev}
\affiliation{Budker Institute of Nuclear Physics, SB RAS, Novosibirsk, 630090, Russia}
\affiliation{Novosibirsk State University, Novosibirsk, 630090, Russia}
\author{K.~I.~Beloborodov}
\affiliation{Budker Institute of Nuclear Physics, SB RAS, Novosibirsk, 630090, Russia}
\affiliation{Novosibirsk State University, Novosibirsk, 630090, Russia}
\author{A.~V.~Berdyugin}
\affiliation{Budker Institute of Nuclear Physics, SB RAS, Novosibirsk, 630090, Russia}
\affiliation{Novosibirsk State University, Novosibirsk, 630090, Russia}
\author{A.~G.~Bogdanchikov}
\affiliation{Budker Institute of Nuclear Physics, SB RAS, Novosibirsk, 630090, Russia}
\author{A.~A.~Botov}
\affiliation{Budker Institute of Nuclear Physics, SB RAS, Novosibirsk, 630090, Russia}
\author{D.~E.~Chistyakov}
\affiliation{Budker Institute of Nuclear Physics, SB RAS, Novosibirsk, 630090, Russia}
\affiliation{Novosibirsk State University, Novosibirsk, 630090, Russia}
\author{T.~V.~Dimova}
\affiliation{Budker Institute of Nuclear Physics, SB RAS, Novosibirsk, 630090, Russia}
\affiliation{Novosibirsk State University, Novosibirsk, 630090, Russia}
\author{V.~P.~Druzhinin}
\email{druzhinin@inp.nsk.su}
\affiliation{Budker Institute of Nuclear Physics, SB RAS, Novosibirsk, 630090, Russia}
\affiliation{Novosibirsk State University, Novosibirsk, 630090, Russia}
\author{L.~V.~Kardapoltsev}
\affiliation{Budker Institute of Nuclear Physics, SB RAS, Novosibirsk, 630090, Russia}
\affiliation{Novosibirsk State University, Novosibirsk, 630090, Russia}
\author{A.~S.~Kasaev}
\affiliation{Budker Institute of Nuclear Physics, SB RAS, Novosibirsk, 630090, Russia}
\author{A.~A.~Kattsin}
\affiliation{Budker Institute of Nuclear Physics, SB RAS, Novosibirsk, 630090, Russia}
\author{A.~G.~Kharlamov}
\affiliation{Budker Institute of Nuclear Physics, SB RAS, Novosibirsk, 630090, Russia}
\affiliation{Novosibirsk State University, Novosibirsk, 630090, Russia}
\author{I.~A.~Koop}
\affiliation{Budker Institute of Nuclear Physics, SB RAS, Novosibirsk, 630090, Russia}
\affiliation{Novosibirsk State University, Novosibirsk, 630090, Russia}
\author{A.~A.~Korol}
\affiliation{Budker Institute of Nuclear Physics, SB RAS, Novosibirsk, 630090, Russia}
\affiliation{Novosibirsk State University, Novosibirsk, 630090, Russia}
\author{D.~P.~Kovrizhin}
\affiliation{Budker Institute of Nuclear Physics, SB RAS, Novosibirsk, 630090, Russia}
\author{A.~S.~Kupich}
\affiliation{Budker Institute of Nuclear Physics, SB RAS, Novosibirsk, 630090, Russia}
\affiliation{Novosibirsk State University, Novosibirsk, 630090, Russia}
\author{A.~P.~Kryukov}
\affiliation{Budker Institute of Nuclear Physics, SB RAS, Novosibirsk, 630090, Russia}
\author{N.~A.~Melnikova}
\affiliation{Budker Institute of Nuclear Physics, SB RAS, Novosibirsk, 630090, Russia}
\affiliation{Novosibirsk State University, Novosibirsk, 630090, Russia}
\author{N.~Yu.~Muchnoi}
\affiliation{Budker Institute of Nuclear Physics, SB RAS, Novosibirsk, 630090, Russia}
\affiliation{Novosibirsk State University, Novosibirsk, 630090, Russia}
\author{A.~E.~Obrazovsky}
\affiliation{Budker Institute of Nuclear Physics, SB RAS, Novosibirsk, 630090, Russia}
\author{E.~V.~Pakhtusova}
\affiliation{Budker Institute of Nuclear Physics, SB RAS, Novosibirsk, 630090, Russia}
\author{K.~V.~Pugachev}
\affiliation{Budker Institute of Nuclear Physics, SB RAS, Novosibirsk, 630090, Russia}
\affiliation{Novosibirsk State University, Novosibirsk, 630090, Russia}
\author{S.~A.~Rastigeev}
\affiliation{Budker Institute of Nuclear Physics, SB RAS, Novosibirsk, 630090, Russia}
\author{Yu.~A.~Rogovsky}
\affiliation{Budker Institute of Nuclear Physics, SB RAS, Novosibirsk, 630090, Russia}
\affiliation{Novosibirsk State University, Novosibirsk, 630090, Russia}
\author{A.~I.~Senchenko}
\affiliation{Budker Institute of Nuclear Physics, SB RAS, Novosibirsk, 630090, Russia}
\author{S.~I.~Serednyakov}
\affiliation{Budker Institute of Nuclear Physics, SB RAS, Novosibirsk, 630090, Russia}
\affiliation{Novosibirsk State University, Novosibirsk, 630090, Russia}
\author{Yu.~M.~Shatunov}
\affiliation{Budker Institute of Nuclear Physics, SB RAS, Novosibirsk, 630090, Russia}
\author{S.~P.~Sherstyuk}
\affiliation{Budker Institute of Nuclear Physics, SB RAS, Novosibirsk, 630090, Russia}
\affiliation{Novosibirsk State University, Novosibirsk, 630090, Russia}
\author{D.~A.~Shtol}
\affiliation{Budker Institute of Nuclear Physics, SB RAS, Novosibirsk, 630090, Russia}
\author{Z.~K.~Silagadze}
\affiliation{Budker Institute of Nuclear Physics, SB RAS, Novosibirsk, 630090, Russia}
\affiliation{Novosibirsk State University, Novosibirsk, 630090, Russia}
\author{I.~K.~Surin}
\affiliation{Budker Institute of Nuclear Physics, SB RAS, Novosibirsk, 630090, Russia}
\author{Yu.~V.~Usov}
\affiliation{Budker Institute of Nuclear Physics, SB RAS, Novosibirsk, 630090, Russia}
\author{V.~N.~Zhabin}
\affiliation{Budker Institute of Nuclear Physics, SB RAS, Novosibirsk, 630090, Russia}
\affiliation{Novosibirsk State University, Novosibirsk, 630090, Russia}
\author{Yu.~M.~Zharinov}
\affiliation{Budker Institute of Nuclear Physics, SB RAS, Novosibirsk, 630090, Russia}

\collaboration{SND Collaboration}

\maketitle

\section{Introduction\label{intro}}
This work is devoted to the measurement of the cross section for the
$e^+e^-\to K_SK_L$ process in the  center-of-mass energy region 
$E=\sqrt{s}=1000$--1100 MeV with the SND detector at the VEPP-2000 collider.
In this region, the dominant contribution to the cross section comes from
the $\phi(1020)$ resonance. The non-resonant contribution coming from the
$\rho$ and $\omega$ tails and excited vector resonances is less than 0.1\% at
the resonance maximum~\cite{cmd-3kskl}. Therefore, the $e^+e^-\to K_S K_L$
process is the best for determining the mass and width of the $\phi(1020)$
resonance. In particular, the values of these parameters obtained in this work
are planned to be used in future SND analyses for the processes 
$e^+e^-\to \pi^+\pi^-\pi^0$ and $e^+e^-\to \pi^0\gamma$, in which the effects
of interference with non-resonant amplitudes are significantly larger than in
the process $e^+e^-\to K_SK_L$.

The energy region under study is near the threshold of the $e^+e^-\to
K_SK_L$ reaction. Therefore, its cross section is sensitive to the effects of
final-state interaction (FSI). The FSI contribution increases as the
threshold is approached. For example, at $E=1000$ MeV, the cross section due
to FSI is expected to increase by 16\%~\cite{milsal}. 
In this analysis, we will fit the measured $e^+e^-\to K_SK_L$ cross section
taking into account the contribution of the final-state interaction.

The main task of experiments at the VEPP-2000 $e^+e^-$ collider is the
precision measurement of the total hadronic  cross section, needed, in
particular, to calculate the hadron contribution to the anomalous magnetic
moment of the muon $a_\mu$. To reach the accuracy of the $a_\mu$ calculation
equal to the expected accuracy of the Fermilab experiment (0.14
ppm)~\cite{g-2-tdr}, the cross section of the dominant process
$e^+e^-\to\pi^+\pi^-$ must be measured with an accuracy of 0.2\% (see, for
example,~\cite{Davier}). For the process $e^+e^-\to K_SK_L$ near the
$\phi(1020)$ resonance, the required accuracy is about 1\%.

In the energy range under study, the most accurate measurements of the
$e^+e^-\to K_SK_L$ cross sections were carried out in the SND~\cite{sndphi}
and CMD-2~\cite{cmd2phi} experiments at the VEPP-2M collider and in the
CMD-3~\cite{cmd-3kskl} experiment at the VEPP-2000 collider. The systematic
errors in the cross section in these experiments are 3.2\%, 1.7\%, and
1.8\%, respectively. 
In this analysis we achieve a systematic uncertainty of 0.9\%.

\section{Detector and experiment}
The Spherical Neutral Detector (SND) is a general-purpose non-magnetic
detector collecting data at the VEPP-2000 $e^+e^-$ collider~\cite{vepp2000}.
A detailed description of detector subsystems can be found in
Refs.~\cite{SNDdet1,SNDdet2,SNDdet3,SNDdet4}. The parameters of charged
particles are measured using a nine-layer drift chamber and a proportional
chamber with cathode-strip readout located in a common gas volume. The solid
angle of the tracking system is 94\% of $4\pi$. Its azimuthal and polar
angle resolutions are $0.45^\circ$ and $0.8^\circ$, respectively. A system
of aerogel threshold Cherenkov counters is located around the tracking
system. The most important part of the detector for the current analysis is
the three-layer spherical electromagnetic calorimeter consisting of 1640
NaI(Tl) crystals. The solid angle of the calorimeter is 95\% of $4\pi$. Its
energy resolution for photons is $\sigma_{E_\gamma}/E_\gamma= 4.2\%/\sqrt[4]
{E_\gamma(\mbox{GeV})}$, and the angular resolution is about 1.5$^{\circ}$.
The calorimeter is surrounded by a 10 cm thick iron absorber, behind which
there is a muon system consisting of proportional tubes and
scintillation counters.

The process $e^+e^-\to K_SK_L$  is analyzed in the decay mode $K_S\to
\pi^0\pi^0$. At $E=1020$ MeV the $K_S$ meson has a decay length of about 6
mm and decays in 96\% of cases inside the beam pipe with a radius
of 20 mm. The $K_L$ meson, which has a lifetime 570 times longer, is
absorbed in most events due to nuclear interaction in the detector
calorimeter. In this case, it is detected as one or more photons. In 27\%
of events, $K_L$ crosses the calorimeter without interaction. Thus, most
events of the process under study fall into the class of events with four
or more detected photons. 
The $e^+e^-\to K_SK_L$ process is the dominant multiphoton process in the
$\phi(1020)$ resonance region. The only source of the resonant background is
the process $e^+e^-\to \eta\gamma\to 3\pi^0\gamma$, which has a cross
section 25 times smaller at the resonance maximum.

The Monte Carlo generators of signal and background events take into account
radiative corrections~\cite{radcor}. The angular distribution of the hard
photon emitted from the initial state is generated according to
Ref.~\cite{BM}. The interactions of particles produced in $e^+e^-$
annihilation with the detector material are simulated using the GEANT4
software package~\cite{geant}. The analysis of processes with the $K_L$
meson in the final state depends critically on the correctness of its
nuclear interaction simulation. The nuclear interaction cross section in
GEANT4 was modified according to Ref.~\cite{KKpi} and then reduced by 5\% to
correctly reproduce the fraction of events, in which the $K_L$ meson crossed
the calorimeter without interaction.

The simulation takes into account variations of experimental conditions
during data taking, in particular, dead detector channels and beam-induced
background. The beam background leads to the appearance of spurious photons and
charged particles in detected events. To take this effect into account,
the simulation uses special background events recorded during data taking with a
random trigger, which are superimposed on simulated events.

The analysis uses data recorded by SND in 2018 in the energy range
$E=1000$--1100 MeV. The integrated luminosity accumulated at 18 energy
points is about 20~pb$^{-1}$. To study the background, data recorded below
the threshold of the $e^+e^-\to K_SK_L$ reaction are also used, in
particular, about 0.7~pb$^{-1}$ at energies of 984 and 990 MeV.

During data taking, the average beam energy and the energy spread were measured by a
special system  using the Compton back-scattering of laser photons on the
electron beam~\cite{compton}. The systematic uncertainty of the beam energy
determination by this method was estimated in Ref.~\cite{compton} by
comparison with the energy measurement by the resonance depolarization
method at $E_b = 510$ and 460 MeV, where $E_b$ is the beam energy. It is
$\Delta E_b/E_b = 6\times 10^{-5}$ or about 60 keV for the center-of-mass
energy $E=2E_b=1000$ MeV. This uncertainty characterizes the possible shift
of the energy scale. The relative shift between points of the energy scan is
smaller. Energy measurements performed at a given energy point are averaged
with weights proportional to the integrated luminosity. The obtained energy
values and their errors are listed in Table~\ref{tab1}. The error of the
measured energy includes the statistical error and
the uncertainty due to the
beam energy drift during data taking. Also, the values of the average
energy spread $\sigma_E$ and their statistical errors are listed in
Table~\ref{tab1}. The systematic error of $\sigma_E$ does not exceed
a fractional error of 5\%.
\begin{table}
\caption{The center-of-mass energy ($E$), center-of-mass energy 
spread ($\sigma_E$) and integrated luminosity ($IL$). For
luminosity, the first error is statistical, the second is systematic.
\label{tab1}}
\begin{ruledtabular}
\begin{tabular}{cccccc}
$E$,      GeV & $\sigma_E$, keV & $IL$, ${\rm pb}^{-1}$\\
\hline
$1000.280\pm 0.086$ & $249\pm 58$ & $ 601.2\pm 2.3\pm 11.0$ \\
$1001.908\pm 0.030$ & $335\pm  8$ & $ 634.2\pm 2.5\pm  2.7$ \\
$1005.986\pm 0.022$ & $356\pm 11$ & $1680.2\pm 4.1\pm  7.3$ \\
$1009.596\pm 0.016$ & $352\pm 13$ & $ 725.7\pm 2.7\pm  4.3$ \\
$1015.736\pm 0.018$ & $385\pm 13$ & $ 627.9\pm 2.5\pm  4.2$ \\
$1016.800\pm 0.034$ & $351\pm 20$ & $1650.1\pm 4.4\pm  8.4$ \\
$1017.914\pm 0.032$ & $352\pm 18$ & $1257.5\pm 3.7\pm  7.4$ \\
$1019.078\pm 0.016$ & $373\pm  8$ & $2454.7\pm 5.7\pm 14.5$ \\
$1019.940\pm 0.016$ & $397\pm 11$ & $2637.2\pm 6.0\pm 13.4$ \\
$1020.908\pm 0.014$ & $396\pm 13$ & $1426.6\pm 4.1\pm  6.2$ \\
$1022.092\pm 0.014$ & $363\pm 11$ & $1232.5\pm 3.7\pm  7.3$ \\
$1022.932\pm 0.028$ & $369\pm 23$ & $ 820.0\pm 2.9\pm  4.2$ \\
$1027.736\pm 0.024$ & $373\pm 17$ & $ 659.2\pm 2.6\pm  5.0$ \\
$1033.816\pm 0.036$ & $366\pm 16$ & $ 537.4\pm 2.3\pm  4.1$ \\
$1039.788\pm 0.036$ & $423\pm 23$ & $ 585.8\pm 2.5\pm  9.0$ \\
$1049.804\pm 0.046$ & $427\pm 23$ & $ 634.2\pm 2.6\pm  5.4$ \\
$1060.016\pm 0.032$ & $393\pm 35$ & $ 607.1\pm 2.5\pm  3.6$ \\
$1100.020\pm 0.046$ & $447\pm 13$ & $1426.1\pm 4.4\pm  6.2$ \\
\end{tabular}
\end{ruledtabular}
\end{table}
			 
\section{Luminosity measurement\label{lumme}}
The process $e^+e^-\to \gamma\gamma$ is used for luminosity measurement.
Like the process under study, it does not contain charged particles in the
final state. Therefore, some uncertainties in the measurement of the
$e^+e^-\to K_SK_L$ cross section associated with the hardware event
selection and beam-induced
background are canceled out as a result of normalization to the luminosity.
We select events without charged tracks and with at least two photons. The
energies of the two most energetic photons in the event ($E_{1,2}$) must be
greater than 0.3E. The azimuthal ($\varphi_{1,2}$) and polar
($\theta_{1,2}$) angles of these photons must satisfy the following
conditions:
$|\Delta\varphi|=||\varphi_1-\varphi_2|-180^\circ|<15^\circ$,
$|\Delta\theta|=|\theta_1+\theta_2-180^\circ|<25^\circ$ 
and $\theta^\ast=(180^\circ-|\theta_1-\theta_2|)/2>45^\circ$.
The latter condition limits the range of polar angles for photons. Unlike
the angles $\theta_{1,2}$, the average polar angle of two photons
$\theta^\ast$ is practically insensitive to the position of the event vertex
along the beam axis, which has a spread of $\sigma_z\approx 3$~cm.
\begin{figure*}
\centering
\includegraphics[width=0.45\linewidth]{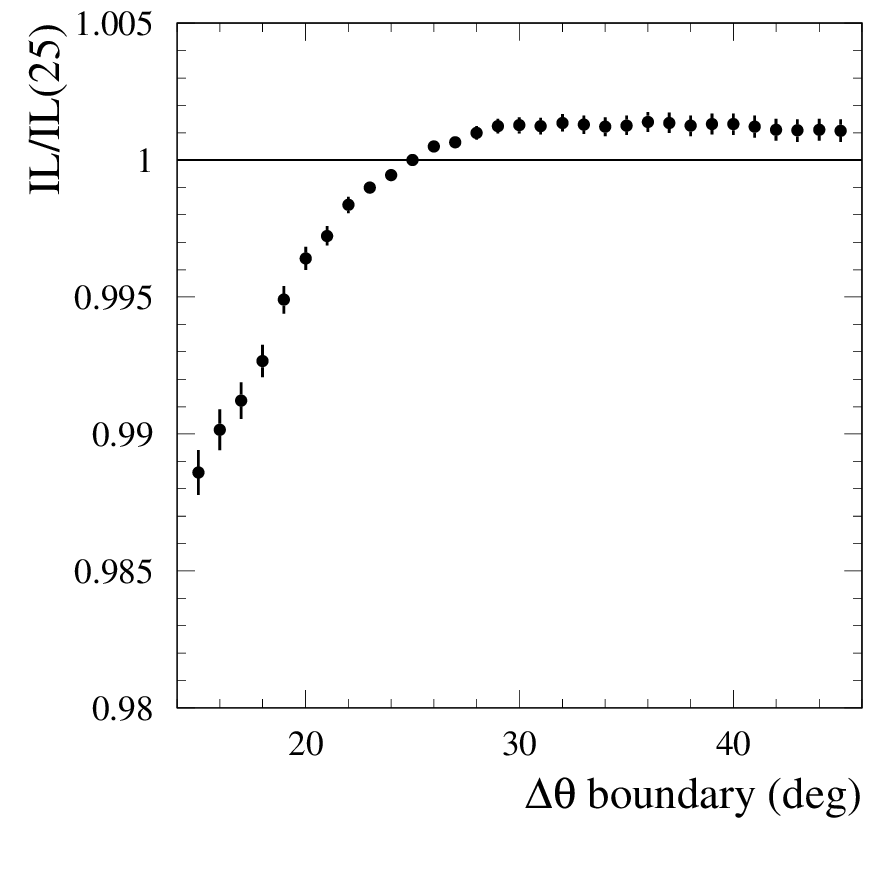}
\includegraphics[width=0.45\linewidth]{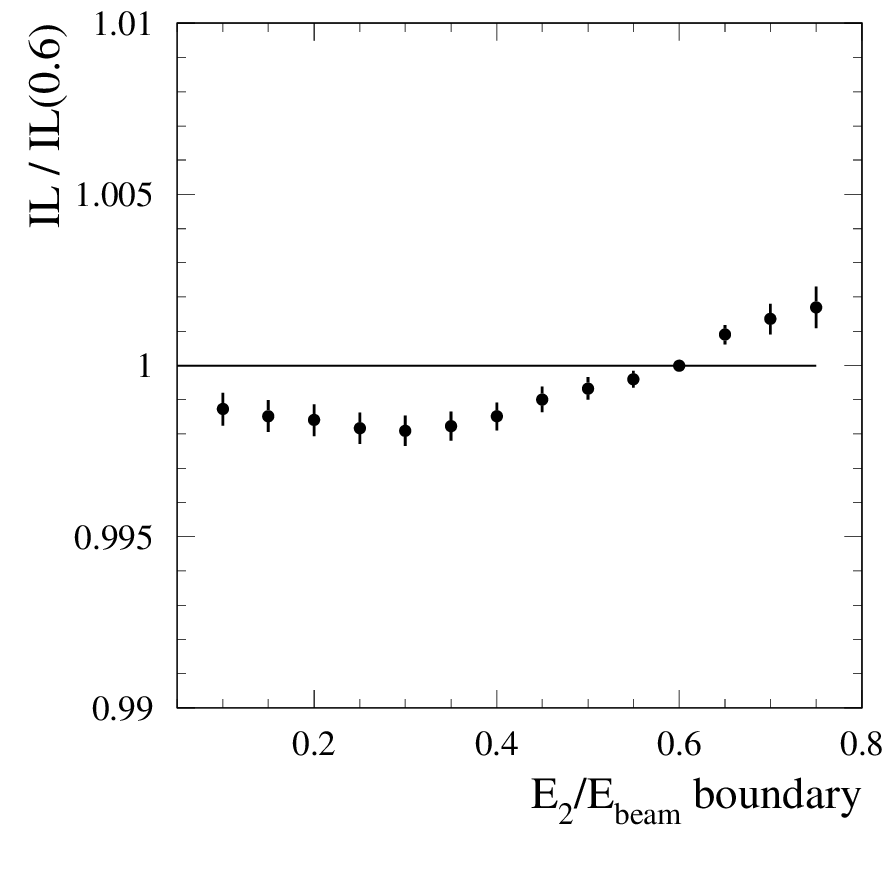}
\caption{
The relative change of the measured integrated luminosity at $E=1019$ MeV
as a function of the boundary on $|\Delta\theta|$ (left) and the photon energy
${\rm min}(E_1,E_2)$ (right).
\label{fig1}}
\end{figure*}
\begin{figure*}
\centering
\includegraphics[width=0.45\linewidth]{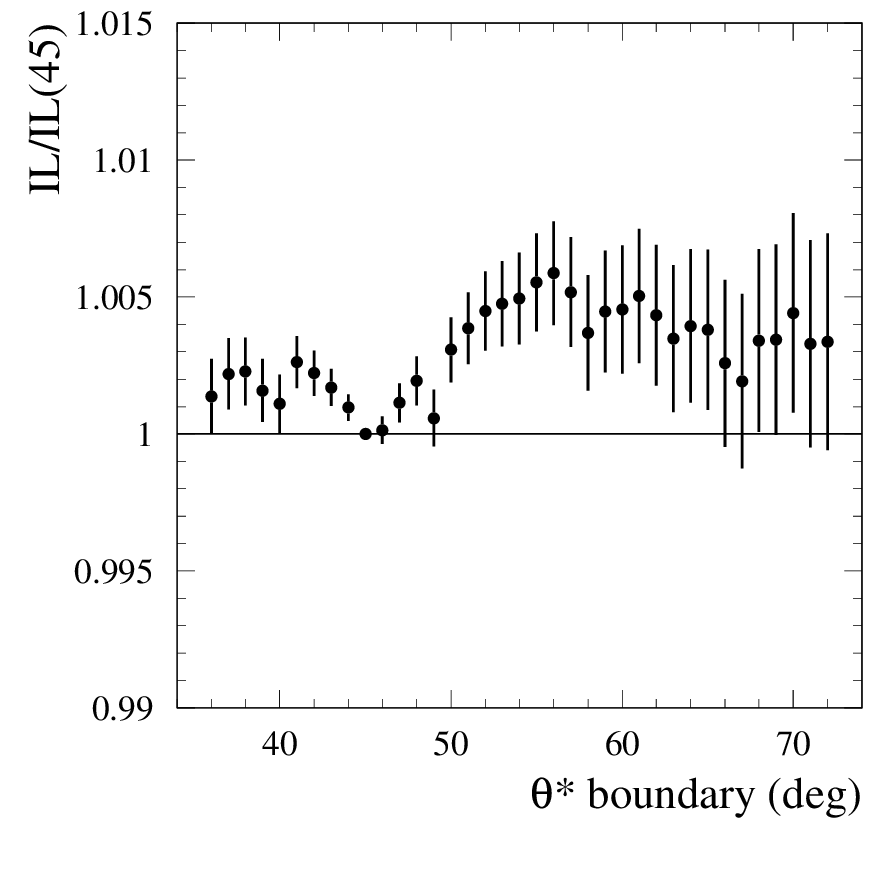}
\includegraphics[width=0.45\linewidth]{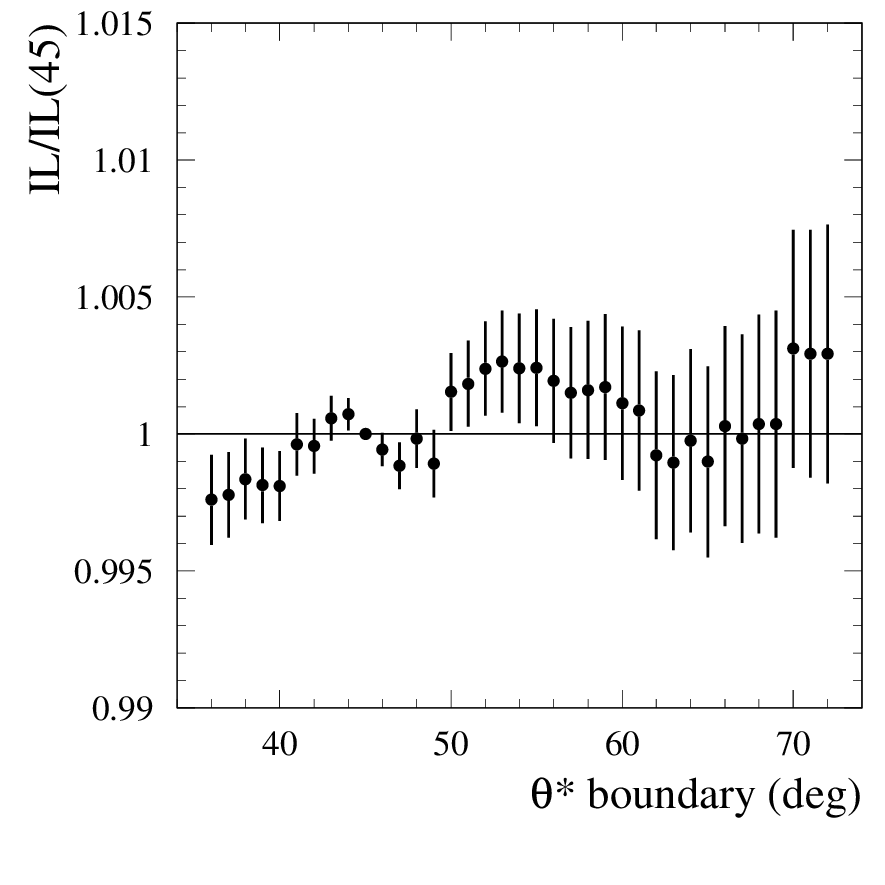}
\caption{The relative change of the measured integrated luminosity
as a function of the boundary on $\theta^\ast$ at $E=1019$ MeV (left)
and 1021 MeV (right).\label{fig2}}
\end{figure*}

The detection efficiency and the cross section for the $e^+e^-\to
\gamma\gamma$ process are determined using the BabaYaga-NLO Monte Carlo event
generator~\cite{babayaga}.  To take into account the inaccuracy of the
detector response simulation for photons, we determine corrections to the
detection efficiency. To do this, the selection conditions described above are
varied in turn over a wide range.  Depending on the presence of background,
the remaining conditions could be tightened. Figure~\ref{fig1} shows how the
measured integrated luminosity at $E=1019$ MeV changes with varying the
conditions on $\Delta\theta$ and the energy of the second least energetic
photon. It is seen that when the conditions on $E_2$ and $\Delta\theta$ are
relaxed, the relative change in luminosity approaches a constant level of about
$-0.15\%$ for $E_2$ and 0.15\% for $\Delta\theta$.
These values with errors of
0.05\% are taken as corrections to the luminosity. The correction for the
condition on $\Delta\varphi$ is not needed. Figure~\ref{fig2} shows the
relative changes in luminosity $IL/IL(45^\circ)$ when varying the condition
on $\theta^\ast$ for $E=1019$ MeV and $E=1021$ MeV. In this
case, no correction is introduced, and the variations of $IL/IL(45^\circ)$
relative to unity are used to estimate the systematic uncertainty in the 
measured luminosity. It is equal to 0.5\% at $E=1019$ MeV and 0.3\% at 
$E=1021$ MeV. Similar studies are performed for all energy points.

Another efficiency correction arises from incorrect simulation of photon
conversion in detector material before the drift chamber. The converted
photon is usually reconstructed as a single charged particle. Such an event
is rejected by the selection conditions. The conversion probability is
measured using events of the process $e^+e^-\to \pi^0\gamma$ at the maximum
of the $\omega(782)$ resonance, where this process has a large cross section
and can be easily separated from background in the class of events with a
single charged particle. The photon conversion probability in data is found
to be greater than that in the simulation by
$(0.23\pm0.02)\%/\sin{\theta_\gamma}$, where $\theta_\gamma$ is the photon
polar angle. For the process $e^+e^-\to \gamma\gamma$ with the selection
described above, the correction for photon conversion is $(0.53\pm0.04)\%$.

In the $e^+e^-\to K_SK_L$ analysis (see Sec.~\ref{cscorr}), we introduce a
correction for the presence of a charged track in the event. In $e^+e^-\to
\gamma\gamma$ events, the main source of extra charged tracks, in addition
to conversion, is superimposing beam-induced background on the
events of interest. To study it, $e^+e^-\to \eta\gamma\to 3\gamma$
events with charged tracks originating outside the beam
interaction region are analyzed. The probability of an extra track in
in the $e^+e^-\to \eta\gamma$ event is found to be $(0.4\pm0.2)\%$ larger in
data than in simulation. This correction is applied to 
$e^+e^-\to \gamma\gamma$ events.
\begin{figure}
\centering
\includegraphics[width=0.90\linewidth]{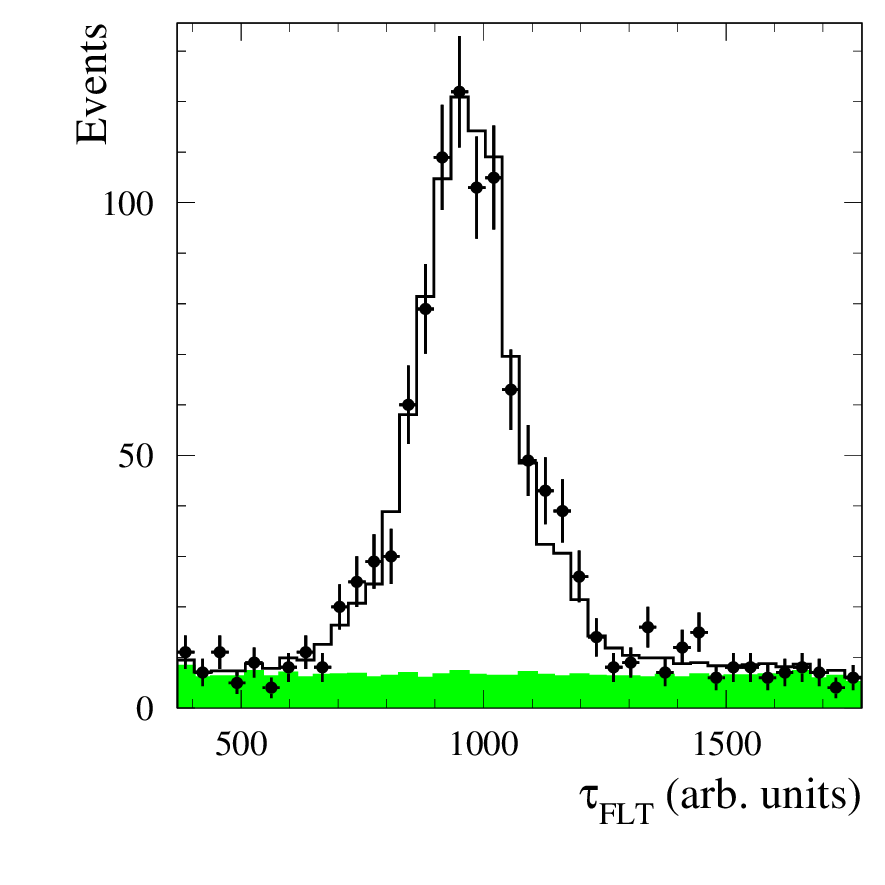}
\caption{
The $\tau_{\rm FLT}$ distribution for $e^+e^-\to \gamma\gamma$
candidate events with $\mu{\rm veto}=1$ at $E=1019$ MeV.
The solid histogram shows the result of the fit by the sum of  distributions
for $e^+e^-\to \gamma\gamma$ events and the cosmic-ray background. The
shaded histogram represents the contribution of cosmic-ray events.
\label{fig3}}
\end{figure}

The background sources for the $e^+e^-\to \gamma\gamma$ events are cosmic
rays and the $e^+e^-\to\pi^0\gamma$ and $e^+e^-\to\eta\gamma$ processes.
Most of the cosmic events that pass the selection conditions for $e^+e^-\to
\gamma\gamma$ have a hit in the muon system ($\mu{\rm veto}=1$). For these
events, we analyze 
the distribution of the arrival time of the calorimeter first-level-trigger
signal relative to the beam collision time ($\tau_{\rm FLT}$).
The
distribution shown in Fig.~\ref{fig3} is fitted by the sum of the peaked
distribution for $e^+e^-\to \gamma\gamma$ events with $\mu{\rm veto}=0$ and 
a flat distribution for cosmic-ray events. The latter is obtained from data
using the special selections described in Sec.~\ref{bkg}.
As a result of the fit, the number of background cosmic-ray
events is obtained. It is about 0.08\% of the total number of $e^+e^-\to
\gamma\gamma$ events and is subtracted. The fraction of cosmic-ray events that
do not fire the muon system ($\mu{\rm veto}=0$) is estimated using data
recorded in 2018 at $E=548$ MeV. The events are processed assuming that 
$E=1020$ MeV. In this case, only background
cosmic-ray events satisfy the $e^+e^-\to \gamma\gamma$ selection criteria.
The fraction of cosmic-ray events with $\mu{\rm veto}=0$ is found to be 20\%.
Thus, the unaccounted cosmic background does not exceed 0.02\%. This number
is used as a measure of the corresponding systematic uncertainty.

The processes $e^+e^-\to\pi^0\gamma$ and $e^+e^-\to\eta\gamma$ imitate
$e^+e^-\to \gamma\gamma$ events when the $\pi^0$ ($\eta$) meson decays along
a direction close to the direction of its motion. The fraction of background
events in the maximum of the $\phi$ resonance determined using simulation is
3\%. The ratio of $e^+e^-\to\pi^0\gamma$ and $e^+e^-\to\eta\gamma$
background events is close to 1:1. The fraction of background events in the
class of events with exactly two photons is 0.3\%. Therefore, the ratio of
the numbers of events with $N_\gamma>2$ and $N_\gamma=2$ can be used to
estimate the accuracy of the background simulation. Its energy dependence
has a resonance component. Its value is reproduced by simulation with an
accuracy of 5\%. Thus, the systematic uncertainty associated with the
resonance background from the processes $e^+e^-\to\pi^0\gamma$ and
$e^+e^-\to\eta\gamma$ does not exceed 0.2\%.
\begin{table}
\caption{The contributions to the systematic uncertainty in the luminosity
measurement ($\Delta IL$) in the energy region 1017--1022 MeV from different
sources.
\label{tab2}}
\begin{ruledtabular}
\begin{tabular}{lc}
Source & $\Delta IL$, \% \\
\hline
Condition $\theta^\ast>45^\circ$    & 0.5 \\
Condition $|\Delta\theta|<25^\circ$ & 0.05 \\
Condition $E_1,E_2>0.3E$            & 0.05 \\ 
Cosmic-ray background               & 0.02 \\
Background from $\phi$ meson decays & 0.2 \\
Photon conversion                   & 0.04 \\   
Beam-induced charged tracks         & 0.2 \\
Theoretical uncertainty             & 0.1 \\
\hline
Total                               & 0.6 \\ 
\end{tabular}
\end{ruledtabular}
\end{table}

The integrated luminosity is determined as follows
\begin{equation}
IL=\frac{N_{\gamma\gamma}-N_{csm}}
{\sigma_{\gamma\gamma}+\sigma_{\eta\gamma}+\sigma_{\pi^0\gamma}},
\end{equation}
where $N_{\gamma\gamma}$ is the number of selected data events,
$N_{csm}$ is the measured cosmic ray background,
$\sigma_{\gamma\gamma}$, $\sigma_{\eta\gamma}$,
and $\sigma_{\pi^0\gamma}$ are the cross sections for the processes
$e^+e^-\to \gamma\gamma$, $e^+e^-\to\eta\gamma$, and $e^+e^-\to\pi^0\gamma$,
respectively,
calculated for the selection criteria described above using simulation.
The distribution of the integrated luminosity over energy points is given in
Table~\ref{tab1} with statistical and systematic errors. For the energy region
1017--1022 MeV, the systematic uncertainty does not exceed 0.6\%. The 
contributions to the systematic uncertainty from different sources are listed
in Table~\ref{tab2}.

\section{Selection of $e^+e^-\to K_SK_L$ events}
\begin{figure*}
\centering
\includegraphics[width=0.45\linewidth]{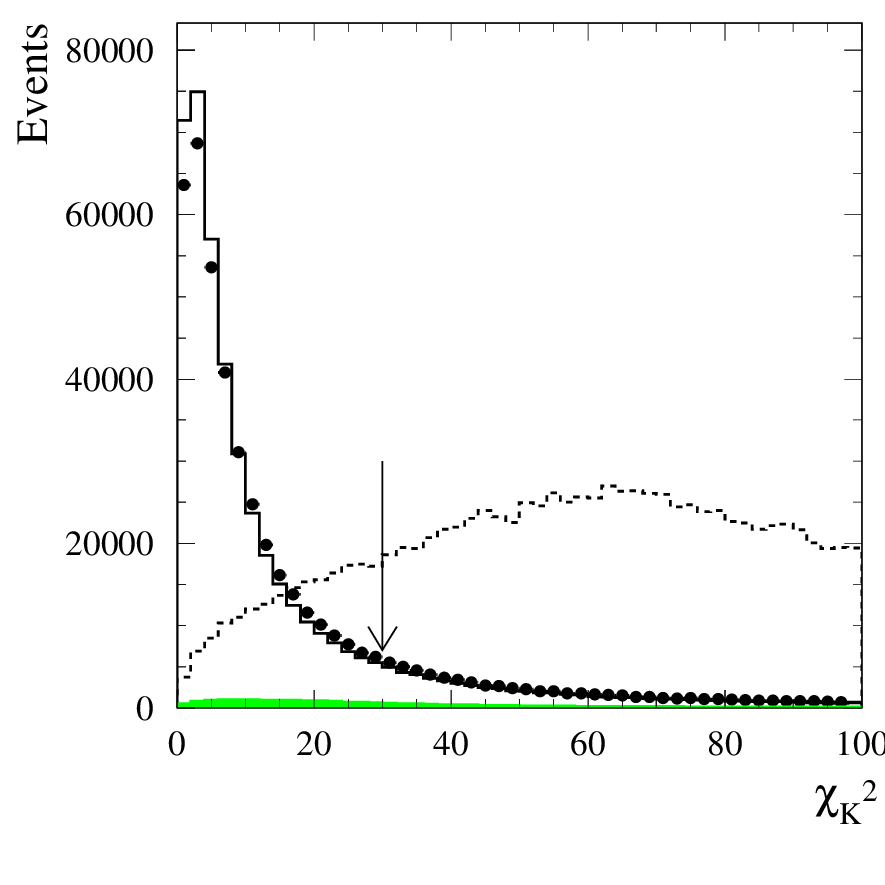}
\includegraphics[width=0.45\linewidth]{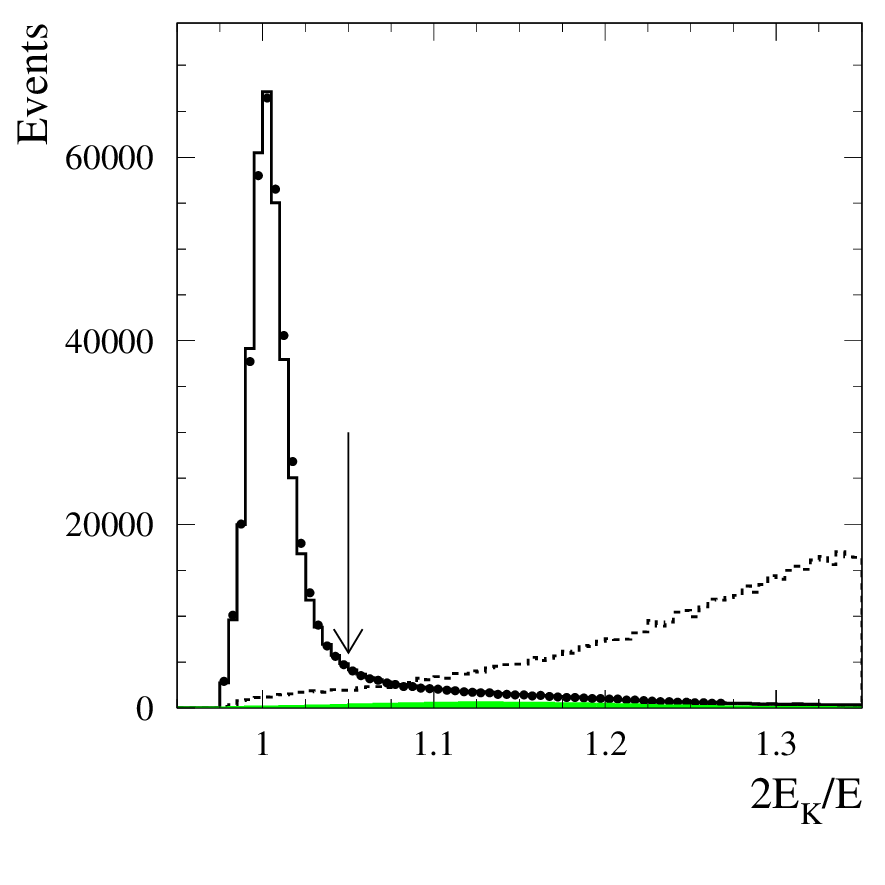}
\caption{
The distributions of $\chi^2_K$ (left) and $2E_K/E$ (right) for data events
(points with error bars) and simulations (solid histogram) at $E=1019$ GeV.
The distributions are normalized to area. The shaded histogram represents
the $e^+e^-\to \eta\gamma$ background calculated using simulation. The
dashed histogram shows the shape of the distribution for the cosmic-ray
background. The arrows indicate the boundaries of the selection conditions.
\label{fig4}}
\end{figure*}

To measure the $e^+e^-\to K_SK_L$ cross section near the threshold , the
selection conditions must provide strong suppression of background events.
Therefore, events are selected in which all four photons from the decay of
$\pi^0$ mesons ($n_\gamma\geq 4)$ are detected. It is required  that there
are no charged tracks in the drift chamber ($n_{\rm ch}=0$). Then, a kinematic
fit of events is performed with the constraints that two pairs of photons form
$\pi^0$ mesons, and two pions form a $K_S$ meson. The quality of the fit is
characterized by the parameter $\chi^2_K$. The distribution of this
parameter for data events at $E=1019$ MeV is shown in Fig.~\ref{fig4}
(left). It is compared with the simulated distribution. The shaded histogram
shows the background contribution from the $e^+e^-\to \eta\gamma$ process
calculated using simulation. All other background sources at this energy
give a negligible contribution. Also shown is the distribution for the
cosmic-ray background obtained from events recorded below the $e^+e^-\to
K_SK_L$  threshold  and selected with the additional requirement of a hit
in the muon system ($\mu\mbox{veto}=1$).

The second parameter used for background suppression is the energy of the
reconstructed $K_S$ meson $E_K$. The $2E_K/E$ distribution is shown in
Fig.~\ref{fig4} (right). To suppress the background, the following
conditions are imposed:
\begin{equation}
\chi^2_K<30,\,2E_K/E<1.05,
\end{equation}
shown in Fig.~\ref{fig4} by arrows.

Events of the $e^+e^-\to K_SK_L$ process can be divided into four classes
listed in Table~\ref{tab3}. The main for analysis Class I contains events
satisfying all the conditions described above. The detection efficiency of
$e^+e^-\to K_SK_L$ events in this class is 50\% at $E=1019$ GeV. 
Events having $n_\gamma \geq 4$ but not satisfying the
conditions $\chi^2_K<30$ and $2E_K/E<1.05$ fall into Class II.
Class III with $n_\gamma\leq 3$ mainly contains events, in which the 
$K_L$ meson does not produce a signal in the calorimeter, and one of the 
photons from the $\pi^0$ decays is lost. This class has a very high level of
beam and cosmic-ray backgrounds.
Class IV contains events with one or more charged tracks. The main
causes of the charged track appearance in $K_SK_L$ events are photon conversion
in material before the drift chamber (4\%), $\pi^0\to e^+e^-\gamma$ decay
(2.3\%), $K_L$ meson decay inside the drift chamber ($\approx 4\%$), and
beam background superimposing on $K_SK_L$ events ($\approx 5\%$). 

To measure the $e^+e^-\to K_SK_L$ cross section and the $\phi$ meson
parameters, the following strategy is used. The background is subtracted
from data events selected with the standard conditions (Class I). 
Then the visible cross section is determined as
\begin{equation}
\sigma_{{\rm vis},i}=\frac{N_{K_SK_L,i}}{\varepsilon_i IL_i},
\label{sigvis}
\end{equation}
where $N_{K_SK_L,i}$, $\varepsilon_i$ and $IL_i$ are the number of selected
$K_SK_L$ events, the detection efficiency calculated using simulation, and
the integrated luminosity, respectively, at the $i$-th energy point. The
measured cross section is fitted with the vector meson dominance (VMD)
model. At this stage, the resonance mass and width are determined, as well
as the relative value of the nonresonant amplitude of the $e^+e^-\to
K_SK_L$ process.
\begin{table}
\caption{Four classes, into which $e^+e^-\to K_SK_L$ events are divided
by the selection conditions, and the fractions of events in these classes
($f$) at $E=1019$ MeV calculated using simulation.
\label{tab3}}
\begin{ruledtabular}
\begin{tabular}{llc}
Class & Selection conditions & $f$, \% \\
\hline
IV  & $n_{\rm ch}>0$                                      & 15.4 \\
III & $(n_{\rm ch}=0)$ {\footnotesize AND} $(n_\gamma <4)$                & 7.1  \\ 
II  & $(n_{\rm ch}=0)$ {\footnotesize AND} $(n_\gamma \geq 4)$
{\footnotesize AND NOT}    & 27.5 \\
    & [$(\chi^2_K<30)$  {\footnotesize AND} $(2E_K/E<1.05)$]&      \\
I   & $(n_{\rm ch}=0)$ {\footnotesize AND} $(n_\gamma \geq 4)$ {\footnotesize AND}    & 50.0 \\
    & $(\chi^2_K<30)$  {\footnotesize AND} $(2E_K/E<1.05)$        &      \\ 
\end{tabular}
\end{ruledtabular}
\end{table}

The obtained model parameters are used in the analysis of Class II events.
From the ratio of the number of $K_SK_L$ events in Classes I and II in the
data and simulation, the correction to the cross section $\delta_{\chi^2}$
is calculated. Then, corrections are obtained for the loss of a photon
$\delta_{\gamma}$ and the presence of a charged track $\delta_{\rm ch}$ in
the event, i.e., for the difference between data and simulation in the
fraction of events falling into Class III and Class IV. 
The methods for determining the corrections
$\delta_{\chi^2}$, $\delta_{\gamma}$, and $\delta_{\rm ch}$ will be described
in detail in Sec.~\ref{cscorr}.
After introducing
the corrections, we obtain the final result for the Born cross section of
the process $e^+e^-\to K_SK_L$ and the parameter $B(\phi\to K_SK_L)B(\phi\to
e^+e^-)$.

\section{Background subtraction \label{bkg}}
\begin{figure*}
\centering
\includegraphics[width=0.45\linewidth]{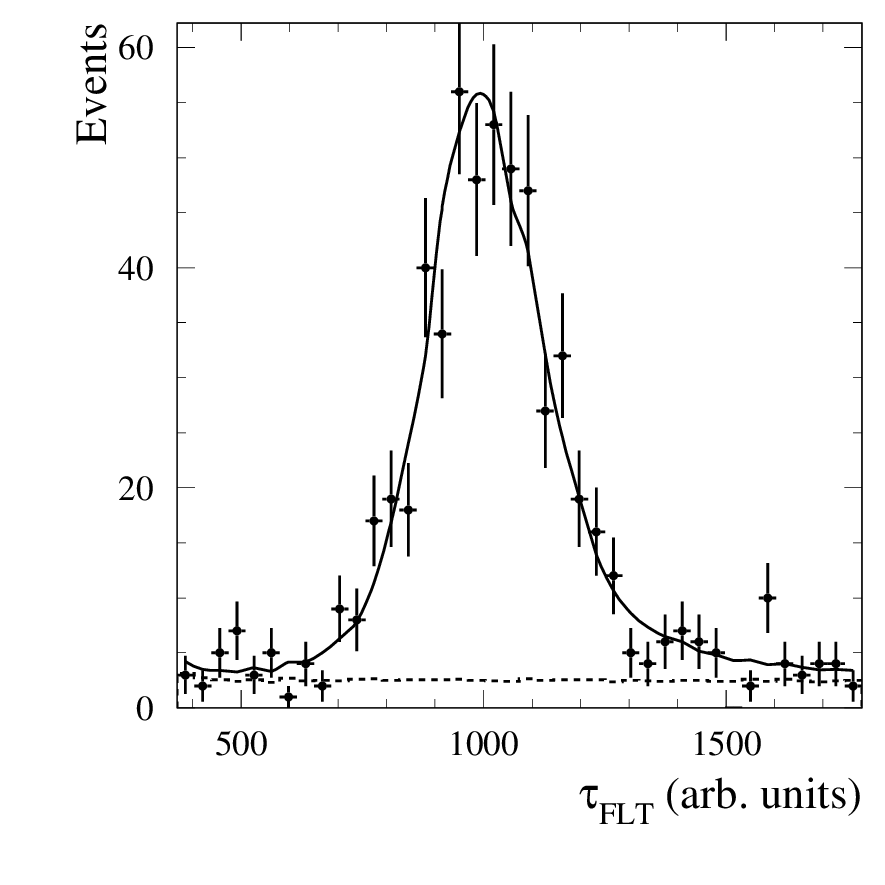}
\includegraphics[width=0.45\linewidth]{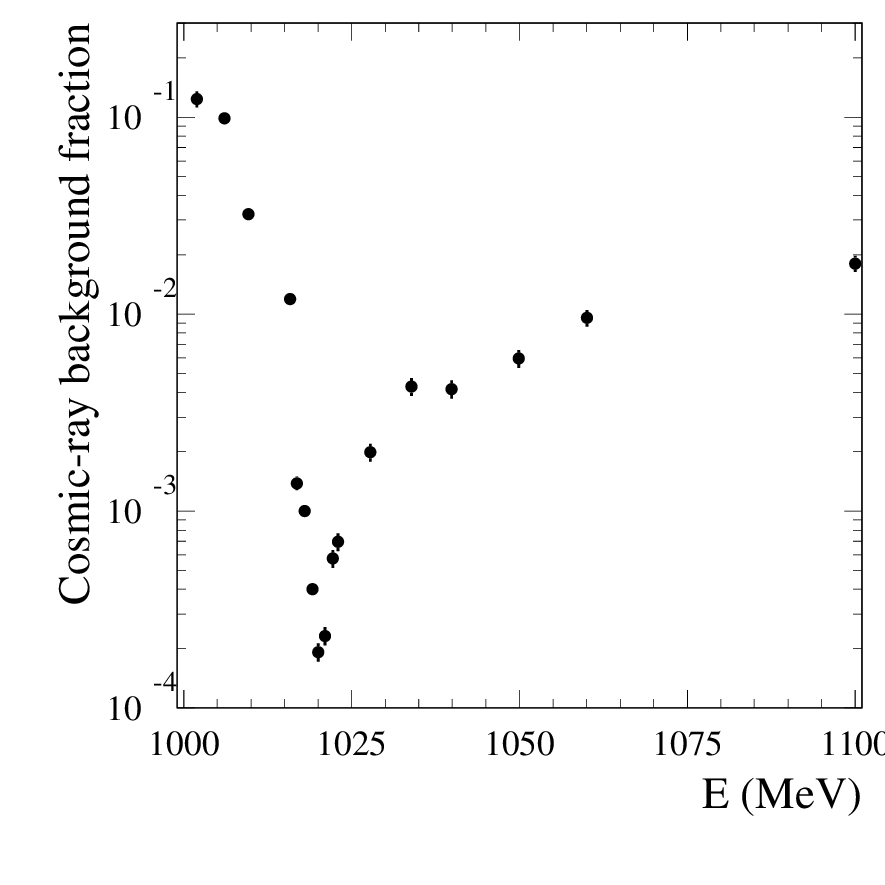}
\caption{
Left panel: The $\tau_{\rm FLT}$ distribution for data events with
$\mu{\rm veto}=1$ at $E=1028$ MeV. The solid curve represents
the result of the fit to this distribution with the sum of signal
and cosmic-ray background. Right panel: The fraction of cosmic-ray background
in Class I.
\label{fig5}}
\end{figure*}
The main sources of background in Class I are the processes
$e^+e^-\to \eta\gamma$ and $e^+e^-\to \pi^0\pi^0\gamma$ and cosmic rays.
Most cosmic-ray events that pass the selection conditions trigger the muon
system ($\mu{\rm veto}=1$). Figure~\ref{fig5} (left) shows the $\tau_{\rm
FLT}$ distribution for events with $\mu{\rm veto}=1$ at $E=1028$ MeV. To
subtract the background, the distribution is fitted by the sum of a peaked
distribution of events $e^+e^-\to K_SK_L$ and a flat distribution for
cosmic rays. The first distribution is obtained using events with $\mu{\rm
veto}=0$ at $E=1019$ MeV, and the second one is obtained using Class II events
with $\mu{\rm veto}=1$ and $30<\chi^2_K<100$, recorded below the $e^+e^-\to
K_SK_L$ threshold at $E=910$--930 MeV.
At $E=1000$ MeV, where the signal from $e^+e^-\to K_SK_L$ events is not
observed in the $\tau_{\rm FLT}$ spectrum due to the relatively high cosmic
background level, events with $\mu{\rm veto}=1$ are rejected. The detection
efficiency at this energy point is corrected accordingly. It should be noted
that the muon system response for $K_SK_L$ events is simulated incorrectly.
At $E>1010$ MeV, the fraction of events with $\mu{\rm veto}=1$ is about 4\%
in the data and 6.5\% in the simulation. At $E\leq 1010$, this fraction
decreases to 5\% in simulation. In data, its average value in the energy
region $E=1002$--1010 MeV is $(4.1\pm 0.6)\%$. Therefore, for the point
$E=1000$ MeV, we increase the detection efficiency by 1\% and introduce an
additional systematic uncertainty also equal to 1\%.

In Class I with $\mu{\rm veto}=0$, the cosmic-ray background is determined
using data with $E=910$--930 MeV. For these data, the $\tau_{\rm FLT}$
distribution is fitted, and the number of cosmic-ray events is
determined. The expected background at the $i\rm{th}$ energy point is calculated as
\begin{equation}
N_{{\rm cosm},i}=N_{{\rm cosm},i}^0\frac{t_i}{t_0},
\label{eqcosm}
\end{equation}
where $t_i$ and $t_0$ are the data taking times at the energy point $i$ and
at $E=910$--930 MeV ($t_0=284000$ s, $t_i\lesssim 100000$ s), and $N_{{\rm
cosm},i}^0$ is the number of cosmic-ray events at $E=910$--930 MeV, obtained
from the fit to the $\tau_{\rm FLT}$ distribution. Due to the condition
$2E_K/E_i<1.05$, the value of $N_{{\rm cosm},i}^0$ increases by a factor of
1.6 when $E_i$ changes from 1000 to 1100 MeV. At the energy points $E=1000$,
1001, 1003 MeV, where the cross section of the process $e^+e^-\to K_SK_L$ is
small, the contribution of the cosmic-ray background for events with
$\mu{\rm veto}=0$ can be determined directly from the fit to the $\tau_{\rm
FLT}$ distribution and then compared with Eq.~(\ref{eqcosm}). Their average
ratio $0.8\pm 0.2$ agrees with unity. The difference of the ratio from unity
by 20\% is used as an estimate of the systematic uncertainty in determining
the cosmic-ray background. The fraction of the beam background in Class I,
shown in Fig.~\ref{fig5} (right), varies from 12\% at $E=1000$ MeV to
$2\times10^{-4}$ at the maximum of the $\phi$ resonance and then to 2\% at
$E=1100$ MeV.
\begin{figure}
\centering
\includegraphics[width=0.9\linewidth]{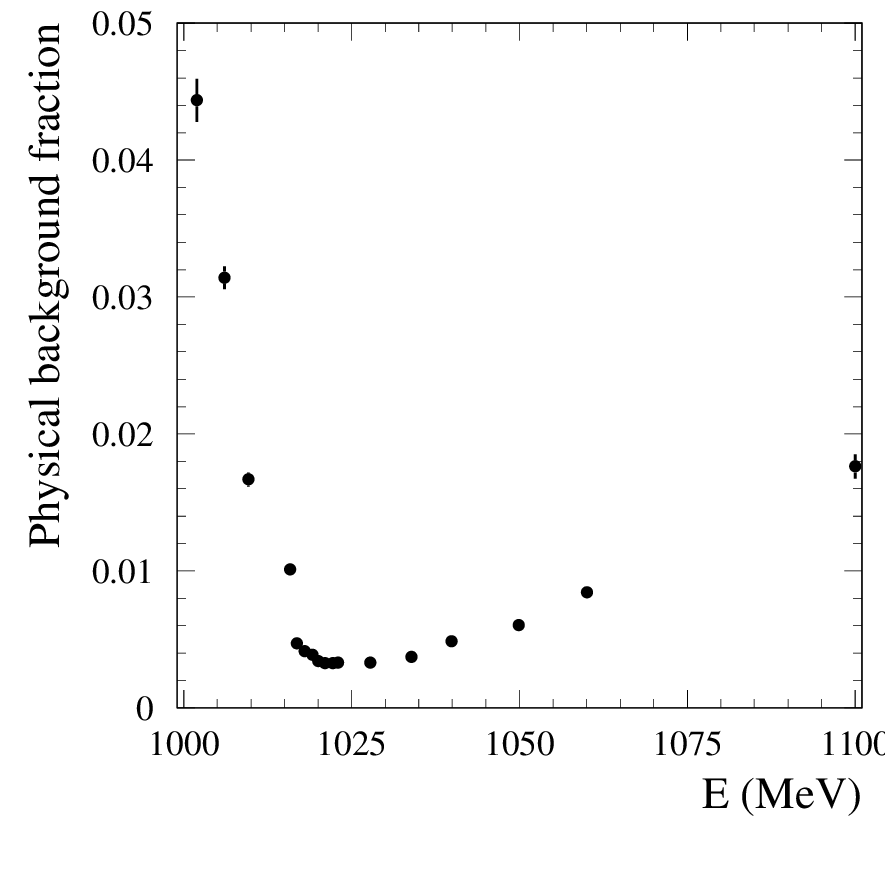}
\caption{The fraction of the  $e^+e^-\to \eta\gamma$ and
$e^+e^-\to\pi^0\pi^0\gamma$ background in Class I.
\label{fig6}}
\end{figure}

The background from the processes $e^+e^-\to \eta\gamma$ and $e^+e^-\to
\pi^0\pi^0\gamma$ is calculated using simulation. The $e^+e^-\to \eta\gamma$
cross section in the event generator is corrected using data as described in
Sec.~\ref{cscorr}. To calculate the $e^+e^-\to \pi^0\pi^0\gamma$ cross
section, the approximation from Ref.~\cite{sndppg} is used. The energy
dependence of the fraction of background events in class I is shown in
Fig.~\ref{fig6}. The dominant contribution to the background comes from the
process $e^+e^-\to \eta\gamma$. The accuracy of the background calculation
is estimated using events from the sideband $30<\chi^2_K<100$, where the
$e^+e^-\to \eta\gamma$ contribution can be measured as described in
Sec.~\ref{cscorr}. It is better than 5\%. The background level changes from
4.4\% at $E=1000$ MeV to $3\times10^{-3}$ at the maximum of the $\phi$
resonance and then to 1.8\% at $E=1100$ MeV.

Data collected near the $e^+e^-\to K_SK_L$ threshold are most sensitive to
the background level. For example, at $E=1000$ MeV, the number of selected
$K_SK_L$ events is $N_{K_SK_L}=128\pm 16$ at $IL=0.60$ pb$^{-1}$. And at two
points with $E=984$ MeV and 990 MeV, which are below the threshold, the
number of selected events before background subtraction is 24, and after
$N_{K_SK_L}=4\pm 8$ at $IL=0.67$ pb$^{-1}$. The data below the
threshold confirm the correctness of the background subtraction procedure.

The numbers of events in Class I after background subtraction are
listed in Table~\ref{tab4}. The first error is statistical,
the second is systematic, related to the uncertainty in background subtraction.
\begin{table*}
\caption{The center-of-mass energy ($E$), detection efficiency
($\varepsilon$), number of selected events ($N_{K_SK_L}$), radiative
correction ($1+\delta$), correction for the energy spread
($1+\delta_E$), Born cross section for the process $e^+e^-\to K_SK_L$
($\sigma$). The first error in the number of events and cross section is
statistical, the second is systematic.
\label{tab4}}
\begin{ruledtabular}
\begin{tabular}{cccccc}
$E$, GeV & $\varepsilon$ & $N_{K_SK_L}$ & $1+\delta$ & $1+\delta_E$ & $\sigma$, nb\\
\hline
1000.280 & 0.129 & $128\pm    16\pm 1 $ &0.710 & 1.003 &$    2.31\pm  0.28\pm  0.05 $ \\
1001.908 & 0.142 & $258\pm    19\pm 5 $ &0.720 & 1.005 &$    3.96\pm  0.29\pm  0.08 $ \\
1005.986 & 0.143 & $2085\pm   49\pm14 $ &0.733 & 1.004 &$   11.76\pm  0.29\pm  0.13 $ \\
1009.596 & 0.145 & $2513\pm   58\pm 6 $ &0.734 & 1.006 &$   32.28\pm  0.79\pm  0.31 $ \\
1015.736 & 0.150 & $20110\pm 170\pm10 $ &0.715 & 1.021 &$  293.02\pm  2.9\pm 2.9 $ \\
1016.800 & 0.151 & $91750\pm 320\pm30 $ &0.709 & 1.019 &$  511.18\pm  2.3\pm 4.6 $ \\
1017.914 & 0.150 & $119590\pm360\pm30 $ &0.707 & 1.008 &$  892.92\pm  4.0\pm 8.3 $ \\
1019.078 & 0.149 & $348570\pm620\pm60 $ &0.721 & 0.973 &$ 1354.54\pm  4.1\pm 12.7 $ \\
1019.940 & 0.148 & $373190\pm640\pm70 $ &0.755 & 0.977 &$ 1297.56\pm  3.8\pm 11.5 $ \\
1020.908 & 0.148 & $157170\pm410\pm30 $ &0.812 & 1.004 &$  916.45\pm  3.8\pm  7.7 $ \\
1022.092 & 0.151 & $92160\pm 320\pm20 $ &0.891 & 1.010 &$  550.02\pm  2.6\pm  5.1 $ \\
1022.932 & 0.151 & $47730\pm 230\pm10 $ &0.947 & 1.009 &$  404.02\pm  2.5\pm  3.6 $ \\
1027.736 & 0.152 & $14050\pm 120\pm10 $ &1.218 & 1.003 &$  114.53\pm  1.4\pm  1.2 $ \\
1033.816 & 0.153 & $5795\pm   80\pm 5 $ &1.459 & 1.001 &$   48.14\pm  1.1\pm  0.5 $ \\
1039.788 & 0.154 & $4298\pm   78\pm 4 $ &1.625 & 1.001 &$   29.39\pm  0.95\pm 0.52 $ \\
1049.804 & 0.151 & $2907\pm   63\pm 4 $ &1.815 & 1.000 &$   16.71\pm  0.71\pm 0.22 $ \\
1060.016 & 0.156 & $2135\pm   49\pm 4 $ &1.941 & 1.000 &$   11.64\pm  0.55\pm 0.17 $ \\
1100.020 & 0.156 & $2262\pm   55\pm 8 $ &2.154 & 1.000 &$    4.73\pm  0.26\pm 0.14 $ \\
\end{tabular}	       
\end{ruledtabular}     
\end{table*}	       

\section{Fitting the measured visible $e^+e^-\to K_SK_L$ cross section.
\label{appr}}
Data on the visible cross section $\sigma_{{\rm vis},i}$ obtained using
Eq.~(\ref{sigvis}) are fitted by the following expression
\begin{eqnarray}
\sigma_{\rm vis}(E) &=& \int \limits_{0}^{x_{max}}
F(E,x) \sigma(E\sqrt{1-x}) dx\nonumber \\
&=&\sigma(E)(1+\delta(E)),
\label{eq2}
\end{eqnarray}
where $x=2E_\gamma/E$, $F(E,x)$ is a function describing the probability of
emission of photons with energy $E_\gamma$ from the initial
state~\cite{radcor}, $\sigma(E)$ is the Born cross section for the
process $e^+e^-\to K_SK_L$. The integration is carried out up to the
kinematic limit $x_{max}=1-4m_{K^0}^2/E^2$.
To take into account the beam energy spread, it is necessary to perform a
convolution of the cross section (\ref{eq2}) with a Gaussian function
describing the energy distribution of events. Since the energy spread is
much smaller than the width of the $\phi$ resonance, we use an approximate
formula instead of convolution:
\begin{eqnarray}
\sigma_{\rm vis}(E) &\Longrightarrow&
\sigma_{\rm vis}(E)+
\frac{1}{2}\frac{d^2\sigma_{\rm vis}}{dE^2}(E)\sigma_E^2\nonumber \\
&=&\sigma_{\rm vis}(E)(1+\delta_E(E))
\label{eq3}
\end{eqnarray}

Near the $\phi$ resonance the uncertainty in the collider energy listed in
Table~\ref{tab1} effectively increases the uncertainty in the measured visible
cross section. In fitting the cross-section energy dependence, the
following term is quadratically added to the statistical error of
$\sigma_{{\rm vis},i}$: 
\begin{equation}
\Delta E_i \frac{d\sigma_{\rm vis}}{dE}(E_i),
\label{desys}
\end{equation}
where $\Delta E_i$ is the uncertainty in the energy of $i$th point.
In the range $E=1016$--1023 MeV, this additional uncertainty is comparable to
or exceeds the statistical error.

To describe the Born cross section, the VMD model is used, which,
in addition to the dominant amplitude of the $\phi$ meson, includes the
amplitudes of the $\rho$ and $\omega$ mesons and an amplitude that takes
into account the contribution of the higher vector resonances~\cite{sndphi}:
\begin{widetext}
\begin{eqnarray}
\sigma(E)&=&\frac{12\pi}{E^3}
\frac{\Gamma(\phi\to K_SK_L)P_K^3(E)}{P_K^3(m_\phi)}
\frac{m_\phi^2}{E^2}
\left |\frac{\sqrt{m_\phi^3\Gamma(\phi\to e^+e^-)}}{D_\phi}e^{i\varphi_\phi}
\right . \nonumber\\
&-&\left. k_{\rm SU3}\left [
\frac{\sqrt{m_\omega^3\Gamma(\omega\to e^+e^-)}}{\sqrt{2}D_\omega}-
\frac{\sqrt{m_\rho^3\Gamma(\rho\to e^+e^-)}}{\sqrt{2}D_\rho}\right ]+
A_0\right |^2,\label{bcs}
\end{eqnarray}
\noindent 
with
\begin{equation}
P_K(E)=\sqrt{E^2/4-m_{K_S}},\,\,D_V=m_V^2-E^2-iE\Gamma_V(E),
\end{equation}
\end{widetext}
where $m_V$, $\Gamma(V\to e^+e^-)$, and $\Gamma_V(E)$ are the mass, partial
width of the decay $V\to e^+e^-$, and the energy-dependent total width of the
vector resonance $V=\rho,\omega,\phi$, $\Gamma(\phi\to K_SK_L)$ is the
partial width of the decay $\phi\to K_SK_L$, $\varphi_\phi$ is the
relative phase of the $\phi$ meson amplitude, $A_0$ is the amplitude
describing the contributions of the higher vector resonances. When
calculating the energy dependence of $\Gamma_V(E)$, decays with
branching fractions greater than 1\% are taken into account. In deriving the
formula~(\ref{bcs}), the SU3 symmetry relationship between the coupling
constants $g_{\rho K_S K_L}=-g_{\omega K_S K_L}=g_{\phi K_S K_L}/\sqrt{2}$
is used. The deviation of the coefficient $k_{\rm SU3}$ from unity
characterizes the SU3 symmetry breaking.

In the fit, the phase $\varphi_\phi$ is set to the value predicted by the
quark model $180^0$, the coefficient $k_{\rm SU3}=1$. Since the energy
region under study is located significantly below that of the excited resonances
of the $\rho$, $\omega$, and $\phi$ families, and their widths decrease 
rapidly with decreasing energy, the amplitude $A_0$ is assumed to be real.
To take into account its small energy dependence, it is parametrized as
$A_0=a_0/(1-E^2/m_{\rho(1450)}^2)$. The free parameters of the fit are the
mass and width of the $\phi$ meson, $a_0$, and the ratio $R_\phi$ of the
product $B(\phi\to K_SK_L)B(\phi\to e^+e^-)$ obtained from our data to its
Particle Data Group (PDG) value~\cite{pdg}. The remaining parameters of 
the model are fixed at their PDG values~\cite{pdg}.
\begin{figure}
\centering
\includegraphics[width=0.95\linewidth]{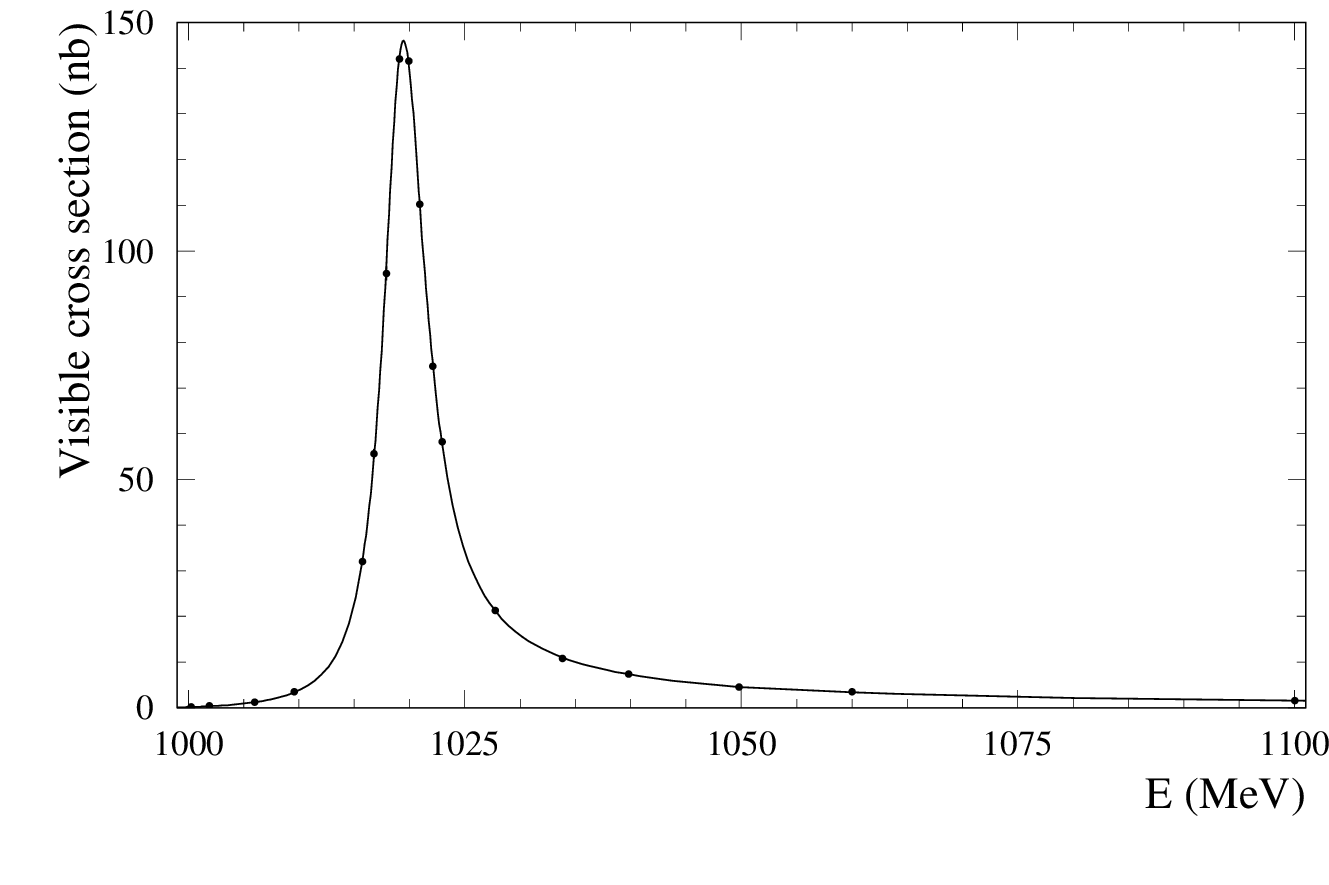}
\caption{The visible $e^+e^-\to K_SK_L$ cross section for Class I events.
The curve is the result of the fit described in the text. 
\label{fig7}}
\end{figure}

The result of the fit to the visible-cross-section data with Eq.~(\ref{eq3})
is shown in Fig.~\ref{fig7}. The quality of the fit is not very good:
$\chi^2/{\rm ndf}=23.6/14$. However, about 8 units of contribution to
$\chi^2$ come from 3 points near the $e^+e^-\to K_SK_L$ threshold. Excluding
these points from the fit leads to an acceptable value $\chi^2/{\rm
ndf}=14.2/11$. In this case, the fit parameters change insignificantly. 
The deviation of the measured cross section from the VMD model
near the threshold can be explained by the final state interaction. A model
taking into account FSI will be discussed in Sec.~\ref{results}. The Born
cross section parameters obtained from the VMD fit will be used in the next
section to obtain corrections.

\section{Corrections to the $e^+e^-\to K_SK_L$ cross section\label{cscorr}}
\begin{figure*}
\centering
\includegraphics[width=0.45\linewidth]{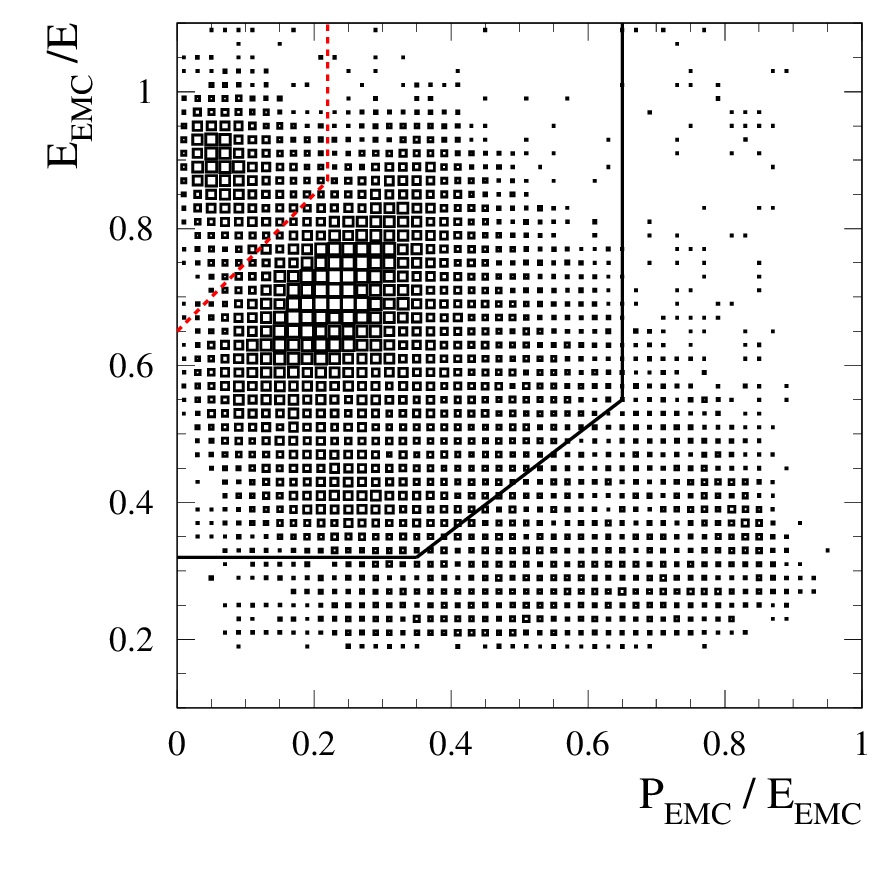}
\includegraphics[width=0.45\linewidth]{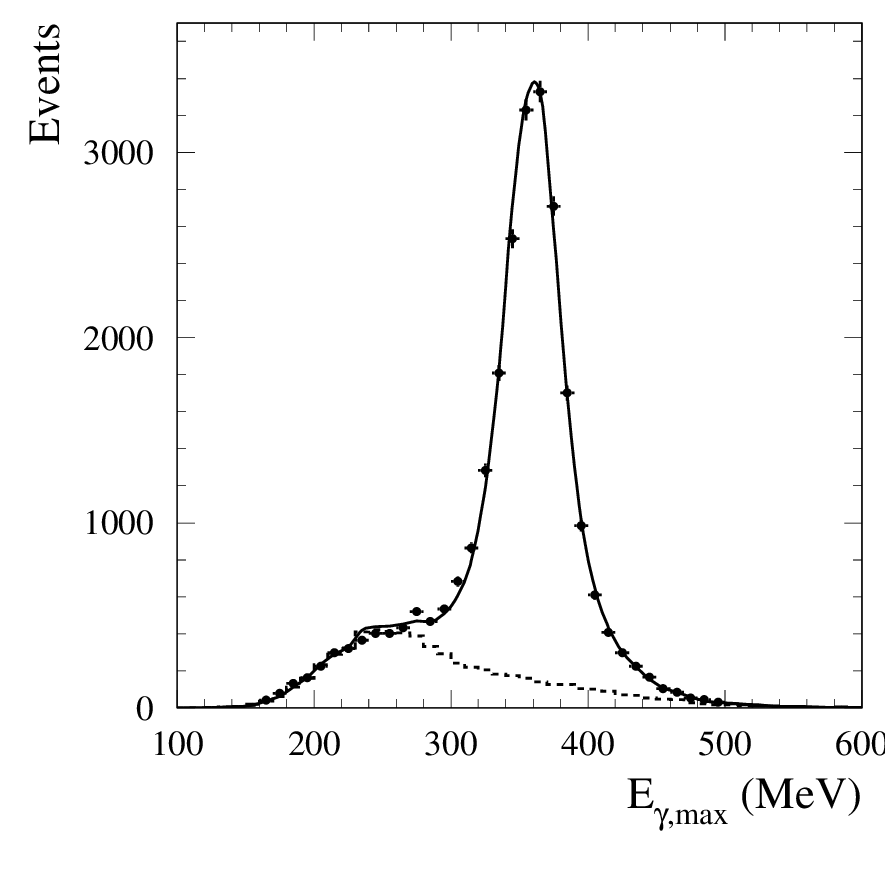}
\caption{Left panel: 
The distribution of $E_{\rm EMC}/E$ versus $P_{\rm EMC}/E_{\rm EMC}$
for Class II data events. The solid broken line indicates the boundary of the
background suppression condition. The dashed broken line indicates
the the boundary of the selection condition for $\eta\gamma$ events.
Right panel: The energy distribution of the most energetic photon
$E_{\gamma,{\rm max}}$ for data events at $E=1019$ MeV. The curve is the
result of the fit by the sum of contributions from processes $e^+e^-\to
\eta\gamma$, $e^+e^-\to K_SK_L$, $e^+e^-\to \pi^0\pi^0\gamma$ and $e^+e^-\to
3\gamma,\,4\gamma$. The dotted curve is the sum of contributions from all
processes except $e^+e^-\to \eta\gamma$. 
\label{fig8}}
\end{figure*}
\begin{figure*}
\centering
\includegraphics[width=0.45\linewidth]{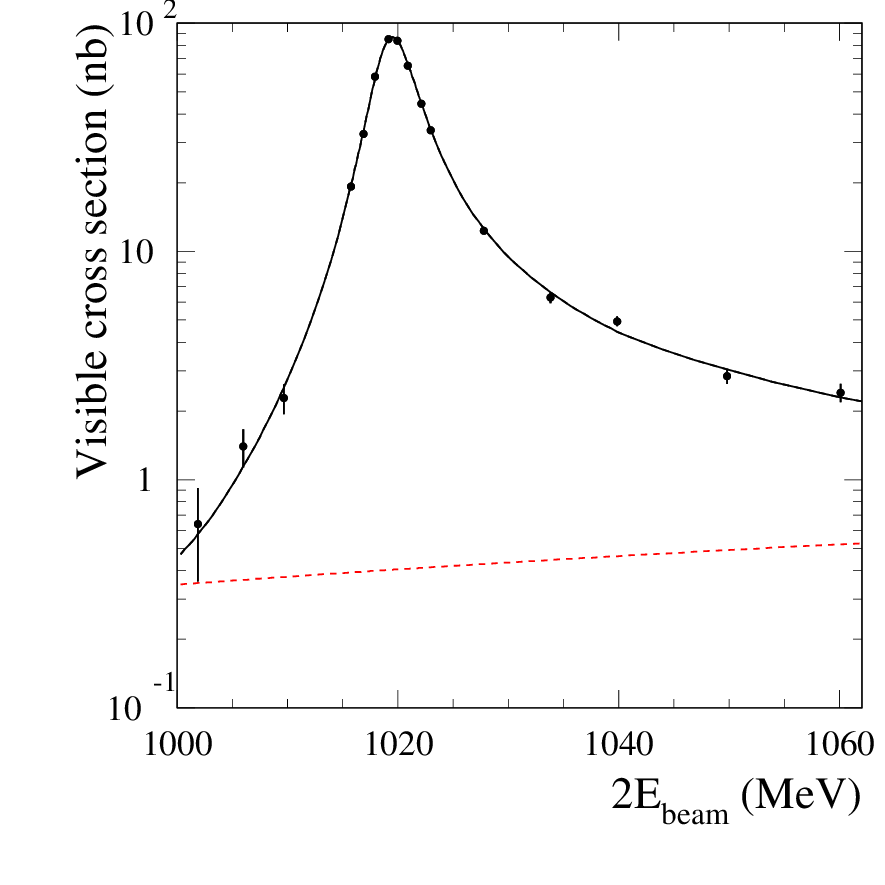}
\includegraphics[width=0.45\linewidth]{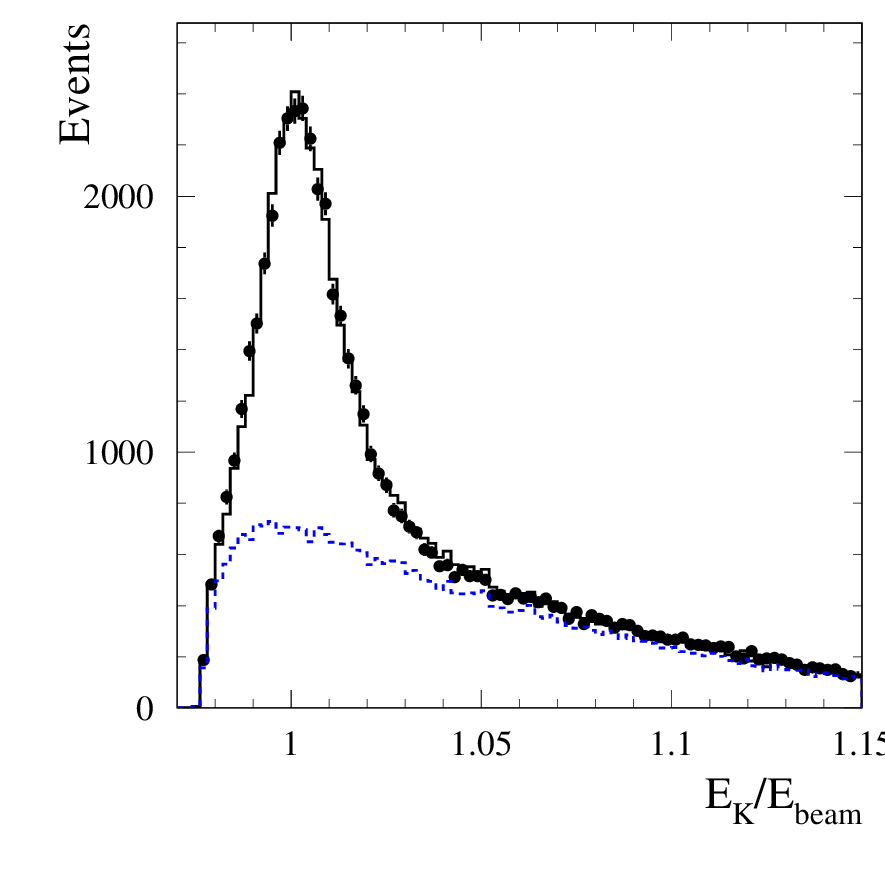}
\caption{Left panel: The visible cross section for events from Class II. The
curve is the result of the fit described in the text. The dotted line shows
the background contribution. Right panel: The $2E_K/E$ distribution for
events with one central charged track at $E=1019$ MeV. The solid histogram
is the result of the fit by the sum of the simulated signal and background
distributions. The dotted histogram is the background distribution.
\label{fig9}}
\end{figure*}
Figure~\ref{fig8} (left) shows the two-dimensional distribution of $E_{\rm
EMC}/E$ versus $P_{\rm EMC}/E_{\rm EMC}$ for data events with four or more
photons that do not satisfy the condition on the parameters $\chi^2_K$ and
$2E_K/E$ (Class II) at $E=1019$ MeV. Here $E_{\rm EMC}$ is the total energy
deposition in the calorimeter, and $P_{\rm EMC}$ is the total event momentum
calculated using the energy depositions in the calorimeter crystals. The
solid
broken line shows the boundary of a selection condition that significantly
suppresses the beam-induced and cosmic-ray backgrounds. Beam background
events usually have a small energy deposition in the calorimeter, and
cosmic-ray background events have a large $P_{\rm EMC}/E_{\rm EMC}$. From
the data recorded below the $e^+e^-\to K_SK_L$ threshold, one can estimate
that this selection condition suppresses the cosmic-ray background by about
10 times, and the beam background is suppressed by 20 times. At the same
time, the number of $K_SK_L$ events in Class II decreases insignificantly,
by 3.6\%. The accumulation of entries in Fig.~\ref{fig8} (left) with $E_{\rm EMC}/E$ near 0.9
and $P_{\rm EMC}/E_{\rm EMC}<0.2$ contains events of the processes
$e^+e^-\to \eta\gamma$ and $e^+e^-\to \pi^0\pi^0\gamma$, as well as the
background from the electrodynamic processes $e^+e^-\to 3\gamma,\,4\gamma$.

Subtraction of remaining cosmic-ray and physical backgrounds is performed as
described in Sec.~\ref{bkg}. The background from the $e^+e^-\to \eta\gamma$
process in Class II is about 12\%. To ensure the required accuracy of
subtraction of this background, we extract $e^+e^-\to \eta\gamma$ events in
the data and determine a scale factor to the number of events expected from
the simulation. To determine the scale factor, we use events with
$N_\gamma>4$, $P_{\rm EMC}/E_{\rm EMC}<0.22$ and $E_{\rm EMC}/E>0.65+P_{\rm
EMC}/E_{\rm EMC}$. The last two conditions are indicated by the dashed
broken line in Fig.~\ref{fig8} (left).
The fraction of $e^+e^-\to \eta\gamma$ events in Class II
satisfying these conditions is 74\%. The energy distribution of the most
energetic photon in an event for data events with $E=1019$ MeV is shown in
Fig.~\ref{fig8}~(right). It is fitted by the sum of the expected
distributions for the simulated events of the processes $e^+e^-\to
\eta\gamma$, $e^+e^-\to K_SK_L$, $e^+e^-\to \pi^0\pi^0\gamma$ and $e^+e^-\to
3\gamma,\,4\gamma$. The free fit parameters are the scale factors for the
expected numbers of $e^+e^-\to \eta\gamma$ and $e^+e^-\to K_SK_L$ events.
The result of the fit is shown in Fig.~\ref{fig8}~(right). The scale factor
for the $e^+e^-\to \eta\gamma$ cross section is determined for all energy
points. At the resonance maximum, its accuracy is 0.8\%. Below 1012 MeV and
above 1026 MeV, average values of the scale factors are used. Their accuracy
is about 4\%. Assuming that the fraction of the remaining
$e^+e^-\to \eta\gamma$ events is reproduced by the simulation with an
accuracy of at least than 10\%, we estimate that the number of background
$e^+e^-\to \eta\gamma$ events is predicted with an accuracy of at least than
3\%.

The visible cross section for Class II events after background subtraction
is fitted by the model described in the previous section with the parameters
fixed at the values obtained in the fit to the cross section for Class I. The
free fit parameters are the scale factor for the Born cross section and two
parameters of the linear function describing the unaccounted background. The
result of the fit is shown in Fig.~\ref{fig9}~(left). It is seen that the
level of the unaccounted background does not exceed 0.5\% at the resonance
maximum. From the fitted value of the scale factor, a correction to the
cross section due to the difference between data and simulation in the
fraction of $K_SK_L$ events rejected by the conditions $\chi^2_K<30$ and
$2E_K/E<1.05$ is calculated. It is found to be $\delta_{\chi^2}=1.001\pm0.004$.
The correction uncertainty is mainly determined by the accuracy of $e^+e^-\to
\eta\gamma$ background subtraction.

The next correction accounts for the difference between data and
simulation in the probability of the $K_SK_L$ event falling into Class III
with $n_\gamma < 4$. It is impossible to analyze events of this class
directly due to the high level of beam and cosmic-ray backgrounds. Instead,
we study events containing a well-identified $K_L$ meson. The $K_L$ candidate
is a particle reconstructed as a single photon with an energy greater than
$0.3E$, which has a transverse distribution of energy depositions in the
calorimeter crystals unlikely for a photon shower.  We compare the results
of the fits to the visible cross sections for events with the $K_L$
candidate and three or four photons. In data, the fraction of events with
three photons turn out to be approximately 13\% larger than in simulation.
The corresponding correction to the $e^+e^-\to K_SK_L$ cross section
is $\delta_\gamma=1.012\pm0.003$.

The last correction is related to $e^+e^-\to K_SK_L$ events containing
charged tracks (Class IV). When analyzing events with $n_{\rm ch}>0$, all
clusters in the calorimeter are considered as photons. The presence of four
or more photons and $\chi^2_K<30$ are required. Events with two central
charged tracks (12\%), one central track (43\%), and no central tracks
(45\%) are analyzed separately. A central charged track is a track
originating from the beam interaction region. The fraction of events with
$n_{\rm ch}>0$ in each subclass is given in the parentheses. The background
sources are the processes $e^+e^-\to K^+K^-$, $e^+e^-\to \pi^+\pi^-\pi^0$
and $e^+e^-\to K_SK_L$ with the decay $K_S\to \pi^+\pi^-$.
To suppress the background in the class with two central tracks, we
additionally require the presence of the $K_L$ candidate, as described in the
previous paragraph. To subtract the background, we fit the $2E_K/E$
distribution by the sum of the simulated signal and background
distributions. The result of the fit for events with one central track at
$E=1019$ MeV is shown in Fig.~\ref{fig9} (right). To estimate the systematic
uncertainty associated with background subtraction, in the classes with zero
and one central track, where the dominant background process is $e^+e^-\to
K^+K^-$, we vary the condition on $\chi^2_K$. In the class with two central
tracks, the distribution of the spatial angle between charged particles is
additionally analyzed, from which information on the relative contributions
of background processes is extracted.
The correction is calculated from the difference between data and simulation
in the ratio of $K_SK_L$ events with $n_{\rm ch}>1$ and $n_{\rm ch}=0$. The
correction to the cross section is $\delta_{\rm ch}=1.014\pm0.005$. The 
correction
uncertainty is determined by the accuracy of background shape simulation.

The total correction to the cross section is $a_{\rm cor}=1.027\pm0.007$.
Technically, this correction is introduced into the detection efficiency
determined from simulation. The corrected detection efficiency
multiplied by $B(K_S\to\pi^0\pi^0)=0.3069\pm0.0005$~\cite{pdg} is listed in
Table~\ref{tab4}. It is seen that the efficiency weakly depends on energy
and is about 15\%.

\section{Results\label{results}}
\begin{table}
\caption{The parameters of the model described in Sec.~\ref{appr} (VMD), and
the same model, but taking into account FSI (VMD+FSI). The number of energy
points used in the fit is given in the parentheses after the model name.
Only statistical errors of the parameters are quoted.
\label{tab5}}
\begin{ruledtabular}
\begin{tabular}{cccc}
Parameter & VMD(18) & VMD(15) & VMD+FSI(18) \\
\hline
$R_\phi$             & $0.974\pm0.003$  & $0.975\pm0.003$  &  $0.975\pm0.003$ \\
$\Delta m_\phi$, MeV & $-0.018\pm0.010$ & $-0.018\pm0.010$ &  $-0.002\pm0.010$ \\
$\Gamma_\phi$, MeV    & $4.230\pm0.019$  & $4.212\pm0.020$  &  $4.212\pm0.019$ \\
$a_0$                & $0.14\pm0.06$    & $0.19\pm0.06$    &  $0.38\pm0.06$ \\
$\chi^2/{\rm ndf}$    & 23.7/14          & 14.2/11          &  13.7/14
\end{tabular}
\end{ruledtabular}
\end{table}
The parameters of the model described in Sec.~\ref{appr}, obtained from the
fit to the visible-cross-section data are listed in Table~\ref{tab5}. The
second column represents the result of the fit to the cross section over all
18 energy points. As already mentioned, 8 units in $\chi^2$ come from the
first three points located near the threshold. The parameters obtained after
removing these points from the fit are listed in the third column. It is
natural to assume that the deviation of the measured cross section from the
model near the threshold can be caused by the interaction of kaons in the
final state. The effect of FSI is taken into account with the
factor~\cite{milsal}, by which the Born cross section (\ref{bcs}) is
multiplied. This factor is normalized to unity at the maximum of the $\phi$
resonance, is equal to 1.16 at $E=1000$ MeV and 0.92 at $E=1100$ MeV. The
fourth column of Table~\ref{tab5} presents the parameters obtained taking
into account FSI. This model describes all the data on the cross section
well. From the difference in $\chi^2$ between the VMD(18) and VMD+FSI(18)
models, we conclude that the significance of the FSI observation in the
process $e^+e^-\to K_SK_L$ near its threshold is $3.2\sigma$.

\begin{figure*}
\centering
\includegraphics[width=0.32\linewidth]{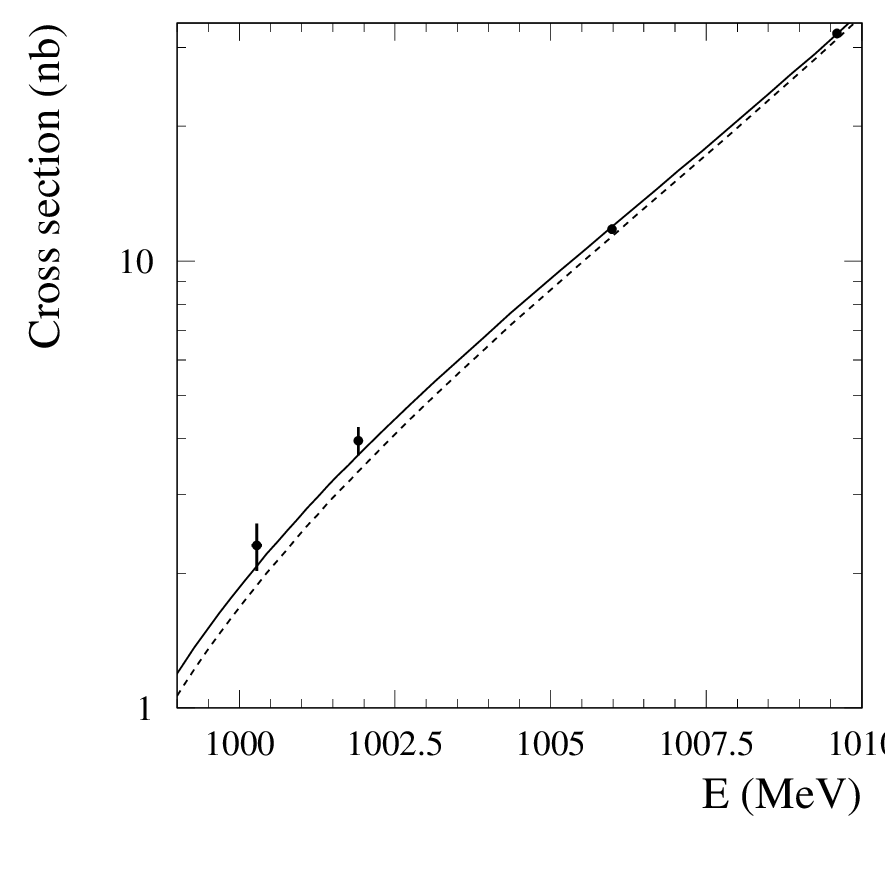}
\includegraphics[width=0.32\linewidth]{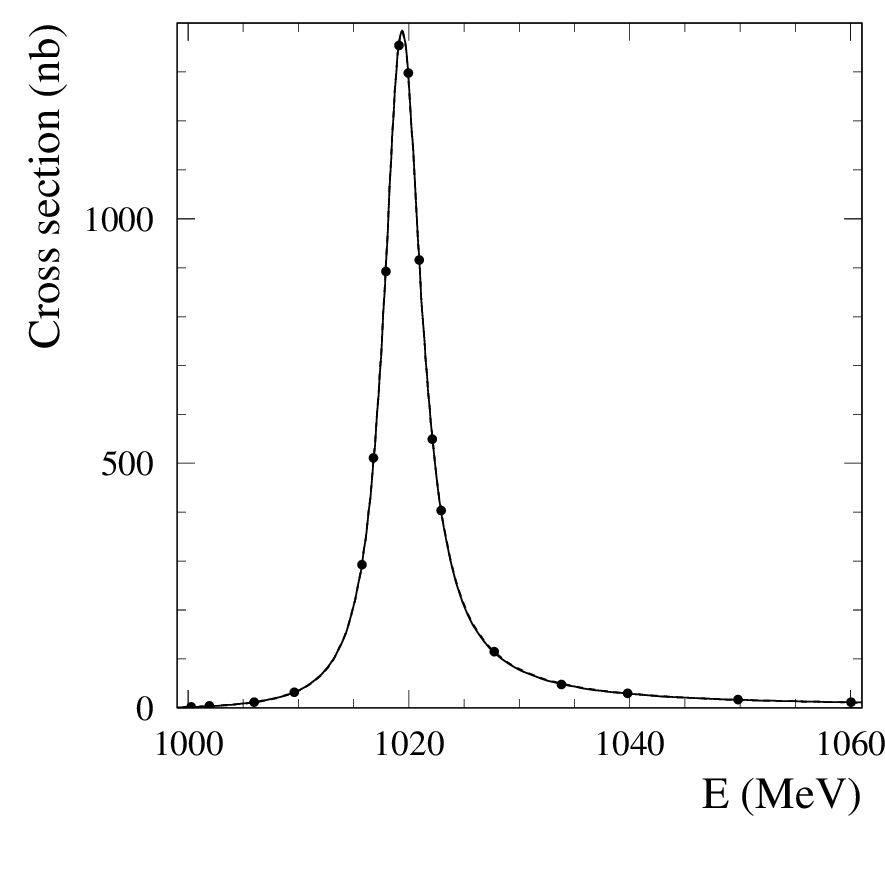}
\includegraphics[width=0.32\linewidth]{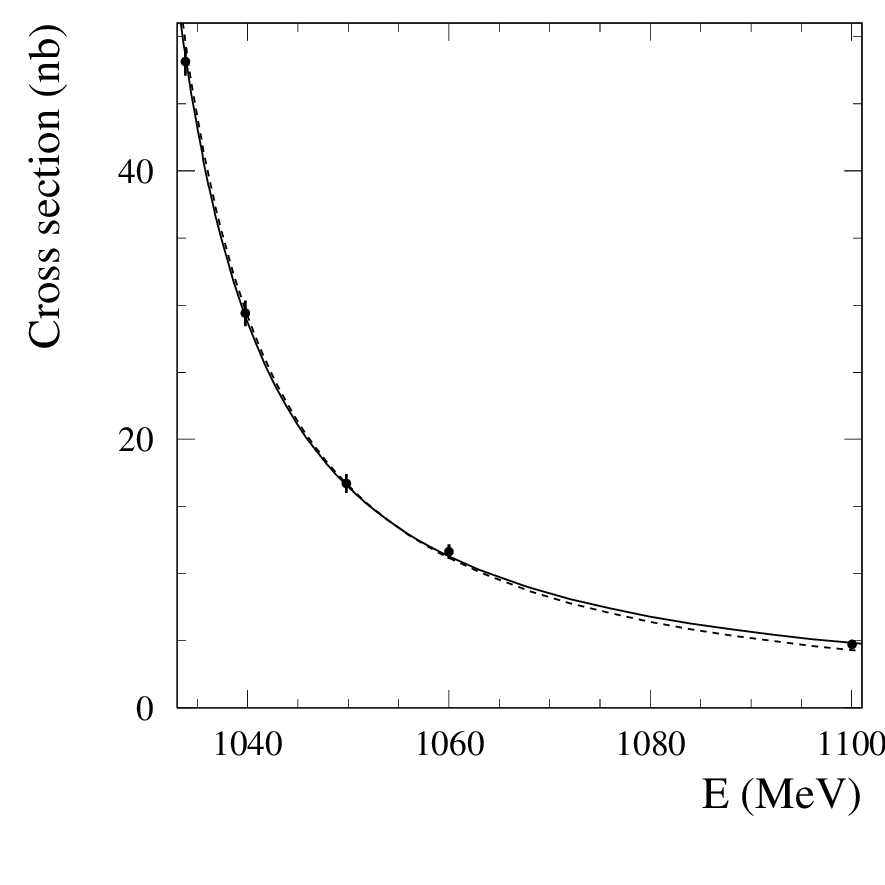}
\caption{The measured Born cross section for the process $e^+e^-\to K_SK_L$. 
The dashed curve is the result of the fit with the VMD(18) model. The solid 
curve represents the VMD+FSI(18) model. 
\label{fig10}}
\end{figure*}
\begin{figure*}
\centering
\includegraphics[width=0.48\linewidth]{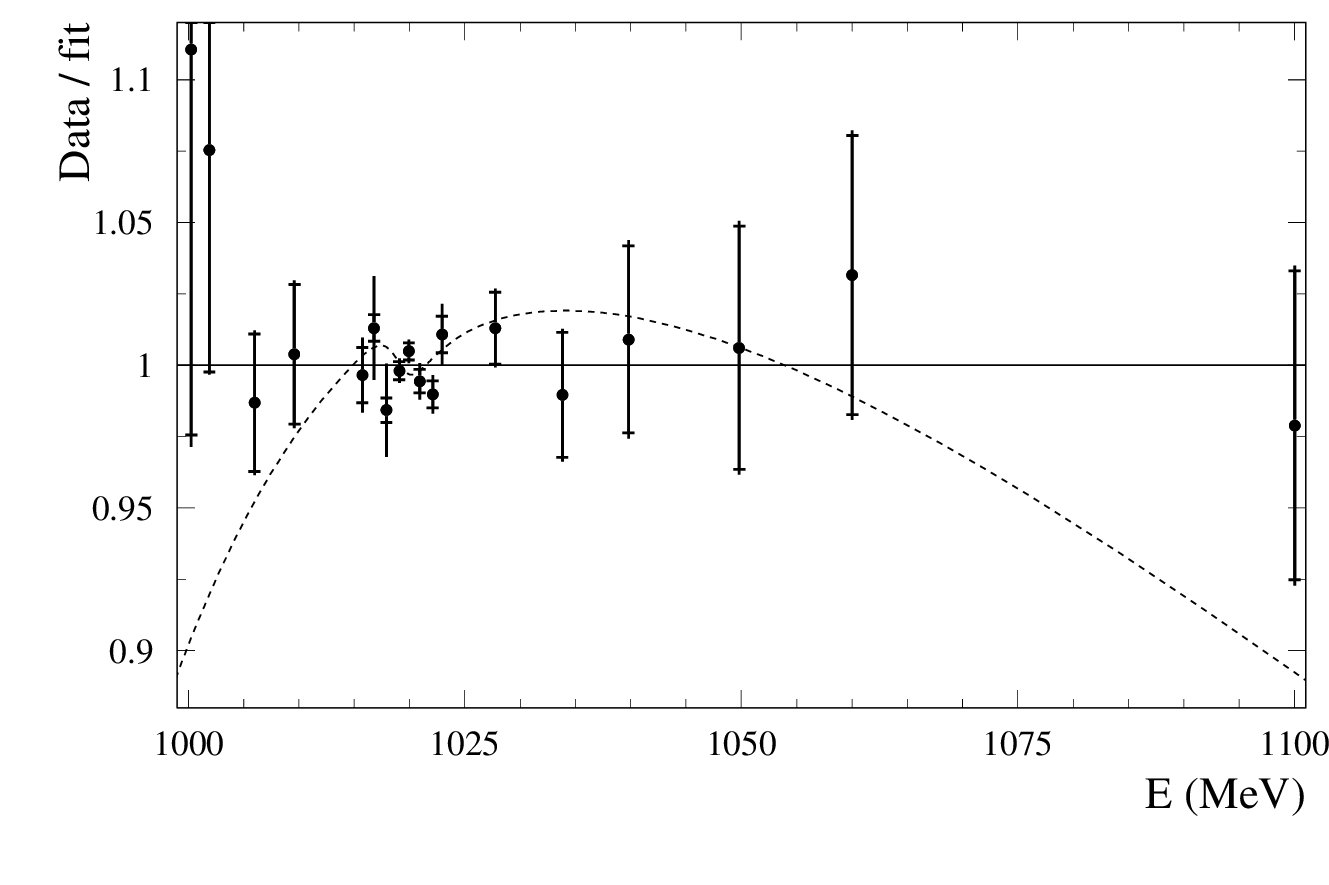}
\includegraphics[width=0.48\linewidth]{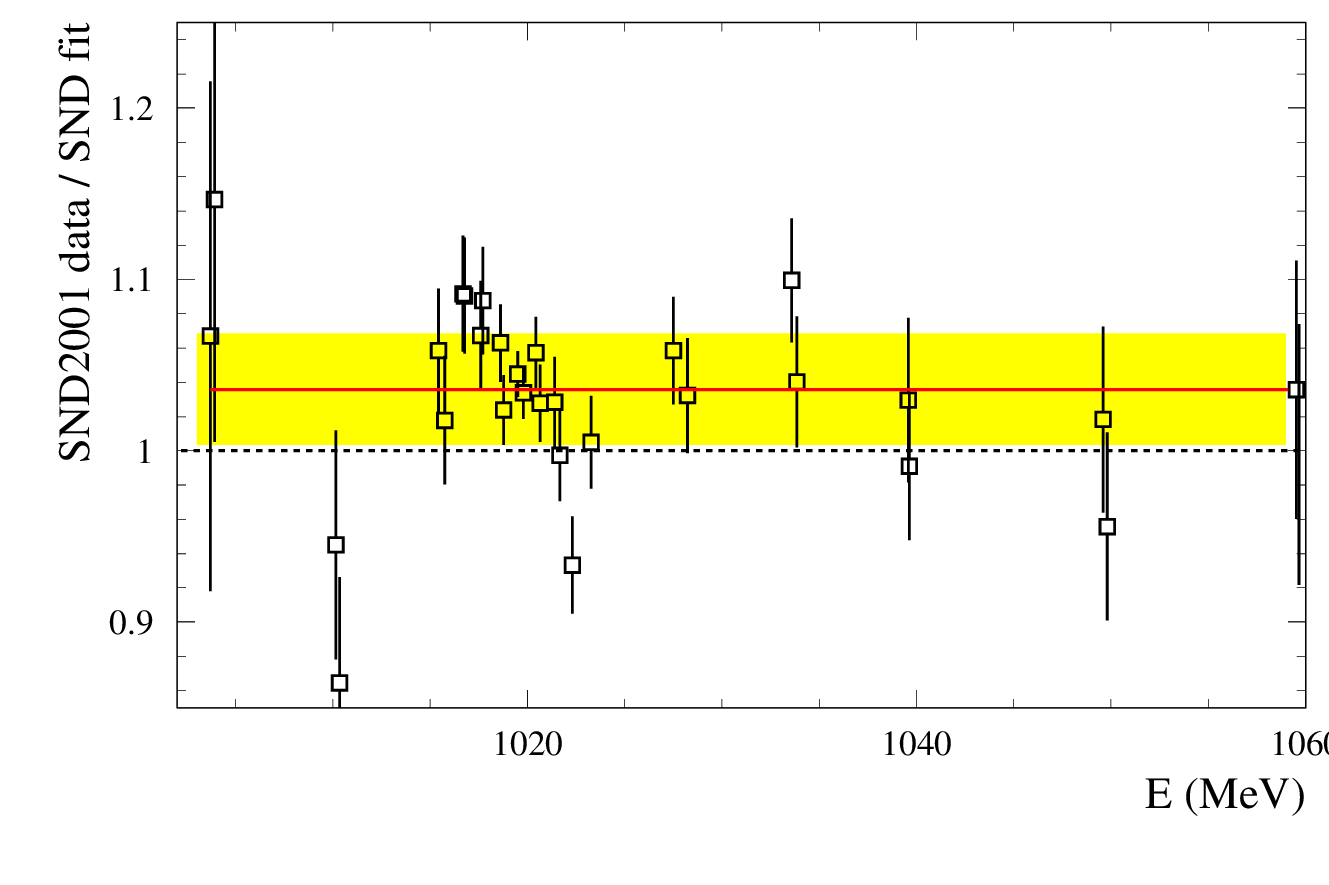}
\caption{Left panel: The ratio of the cross section measured in this work to
the cross section obtained by fitting with the VMD+FSI(18) model. The dashed
curve shows the ratio of the cross sections in the VMD(18) and VMD+FSI(18)
models. For the data points, the the error bars represent the uncertainties
taking into account the inaccuracy of the collider energy setting [see
Eq.~(\ref{desys})], with a tick representing the statistical error only.
Right panel: The ratio of the cross section measured in the SND experiment at
VEPP-2M~\cite{sndphi} to the cross section obtained by fitting the data from
this work with the VMD+FSI(18) model. The solid line represents the result
of the fit to the data with a constant. The shaded box shows the systematic
uncertainty of the SND data.
\label{fig11}}
\end{figure*}

The results of the fit in the VMD+FSI(18) model are used to calculate the
radiative correction and the energy spread correction. After that, the value
of the Born cross section for the process $e^+e^-\to K_SK_L$ at the energy
point $i$ is determined as
\begin{equation}
\sigma_i=\frac{N_{K_SK_L,i}}
{\varepsilon_i IL_i (1+\delta_i)(1+\delta_{E,i})},
\label{eq1}
\end{equation}
The measured energy dependence of the Born cross section is shown in
Fig.~\ref{fig10}. The fitted curves in the VMD(18) and VMD+FSI(18) models
are also presented. The difference between the models is visible in
Fig.~\ref{fig10}~(left), where the region near the $e^+e^-\to K_SK_L$
threshold is shown. This difference is even better visible in
Fig.~\ref{fig11}~(left), where the ratio of $\sigma_i$ to the cross-section
value at the $i$-th point in the VMD+FSI(18) model is
shown. The dotted curve shows the ratio of the cross sections in the VMD(18)
and VMD+FSI(18) models.

The values of $(1+\delta_i)$, $(1+\delta_{E,i})$ and the cross sections with
statistical and systematic errors are listed in
Table~\ref{tab4}. The main contributions to the systematic uncertainty in
the cross section comes from the uncertainties in the luminosity measurement
($\sim0.6\%$) and the correction calculated in Sec.~\ref{cscorr} ($0.7\%$).
The uncertainties associated with the background subtraction, radiative
correction and correction for energy spread are also taken into account. The
latter is due to the uncertainty in the measurement of $\sigma_E$. It is
$0.13\%$ at the resonance maximum. The error in the radiative correction is
determined by varying the model parameters
within their errors. It changes from 0.01\% at $E=1000$ MeV to about 0.1\%
at the resonance maximum and then to 2.8\% at $E=1100$ MeV. At the resonance
maximum, the cross-section accuracy is dominated by a systematic
uncertainty, which is about 0.9\%.
\begin{figure*}
\centering
\includegraphics[width=0.48\linewidth]{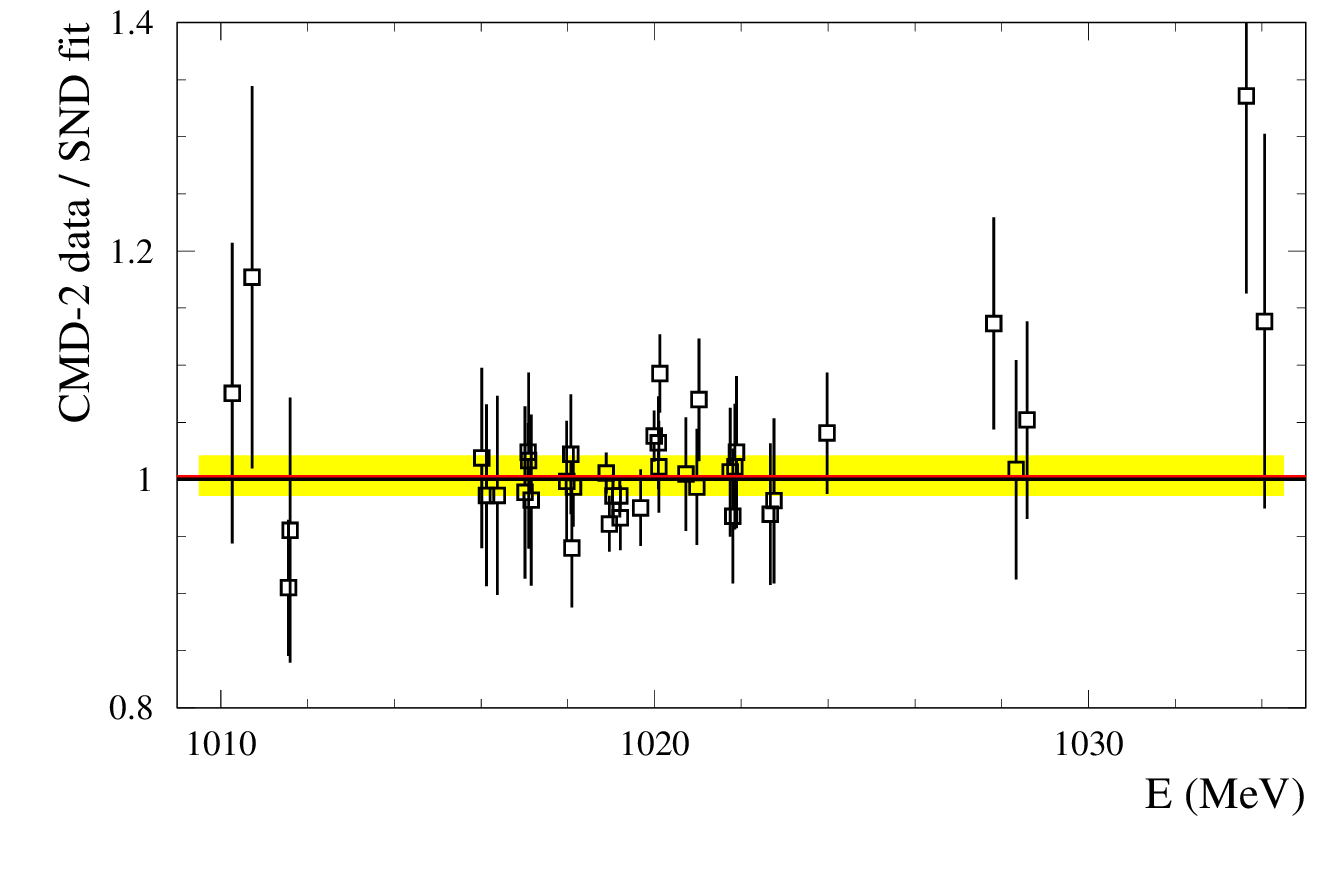}
\includegraphics[width=0.48\linewidth]{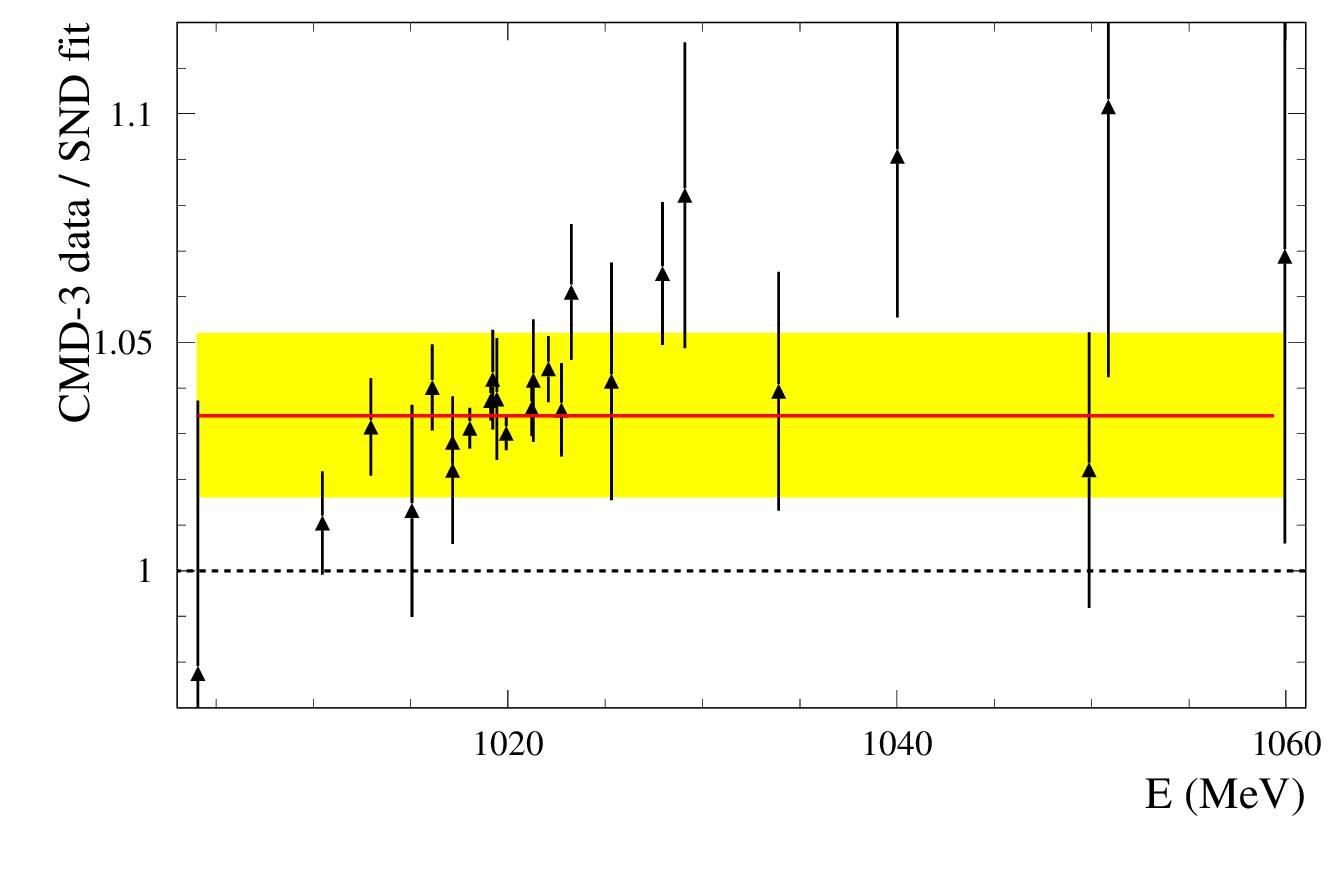}
\caption{The ratios of the cross sections measured in the CMD-2~\cite{cmd2phi}
(left) and CMD-3~\cite{cmd-3kskl} (right) experiments to the cross section 
obtained by fitting the data from this work with the VMD+FSI(18) model.
The solid line represents the result of the fit to the data with a constant. 
The shaded box shows the systematic uncertainty of the data.
\label{fig12}}
\end{figure*}

A comparison of our results with previous measurements is presented in
Figs.~\ref{fig11} and \ref{fig12}, which plot the ratios of the cross
sections measured in Refs.~\cite{sndphi,cmd2phi,cmd-3kskl} to the cross
section obtained by fitting the data from this work in the VMD+FSI(18)
model. The solid line shows the fit of the ratio energy dependence
dependence by a constant. The ratio of the SND measurement at VEPP-2M to the
new measurement is $1.036\pm0.005\pm0.032$. The same ratios for CMD-2 and
CMD-3 are $1.003\pm0.006\pm0.017$ and $1.034\pm0.002\pm0.018$, respectively,
where the first error is statistical and the second is systematic. The
measurements of SND and CMD-2 at VEPP-2M agree with our measurement within
the systematic errors. The difference with the CMD-3 measurement is 1.7
standard deviations.

The VMD(15) model is used to obtain the $\phi$-meson parameters. Similar
models were used to fit the cross-section data in
Refs.~\cite{sndphi,cmd2phi,cmd-3kskl}. As a result of the fit to the
measured cross section, the following values of the $\phi$-meson parameters
are obtained:
\begin{eqnarray}
P_\phi & = & (9.85\pm 0.03 \pm 0.10)\times
10^{-5},\nonumber \\
M_\phi & = & 1019.443\pm 0.010\pm 0.060\mbox{ MeV},\nonumber \\
\Gamma_\phi & = & 4.212\pm0.20\pm0.13\mbox{ MeV},
\end{eqnarray}
where $P_\phi=B(\phi\to e^+e^-)B(\phi\to K_SK_L)$.
The first of the quoted errors are statistical, the second are systematic.
To determine the systematic uncertainties, we shift all $IL_i$ or
$\sigma_{E,i}$ up and down by the value of the systematic error.
The correction $a_{\rm cor}$ is varied within its uncertainty, $\pm 0.7\%$.
The difference between the interference phase $\varphi_\phi$ and the quark 
model prediction ($180^\circ$) can also lead to a change in the parameters. To
estimate the magnitude of the possible phase deviation, we use the result
$\varphi_\phi=(163\pm7)^\circ$ obtained in Ref.~\cite{snd3pi} for the
$e^+e^-\to \pi^+\pi^-\pi^0$ process. To study the systematic uncertainties,
we vary $\varphi_\phi$ from $180^\circ$ to $155^\circ$. The phase of the
amplitude $A_0$ is also changed by $\pm25^\circ$. The parameter $k_{\rm
SU3}$ is varied from 0.9 to 1.1. The systematic uncertainty of the product
$B(\phi\to e^+e^-)B(\phi\to K_SK_L)$ is dominated by the systematic
uncertainties in the luminosity and $a_{\rm cor}$. The uncertainty in the
mass measurement $M_\phi$ is completely determined by the systematic
uncertainty in the collider energy measurement. The main source of the
systematic uncertainty in $\Gamma_\phi$ is the uncertainty in the energy
spread measurement.
 
The measured values of the $\phi$-meson mass and width are consistent with
the PDG values~\cite{pdg}, but are less accurate. The PDG value of the
product $B(\phi\to e^+e^-)B(\phi\to K_SK_L)=(10.11\pm 0.12)\times
10^{-5}$~\cite{pdg} is higher than our measurement by $1.6\sigma$.

The uncertainty of PDG $\phi$-meson mass $1019.461\pm 0.016$ MeV~\cite{pdg}
is significantly smaller than the systematic uncertainty of the collider
energy measurement (60 keV). Therefore, the PDG mass can be used for
calibration of the c.m. energy scale. To do this we introduce into the fit
an additional parameter $\Delta_E$ (common shift of all energy points).
Its fitted value
\begin{equation}
\Delta_E=0.017\pm0.018\mbox{ MeV}
\end{equation}
can be used to correct the energies in the first column of Table~\ref{tab4}.
\section{Summary}
In the SND experiment at the VEPP-2000 collider, the most accurate
measurement of the $e^+e^-\to K_SK_L$ cross section has been performed in
the center-of-mass energy range from 1000 to 1100 MeV. The systematic
uncertainty of this measurement at the maximum of the $\phi$ resonance is
0.9\%. From the fit to the cross section data with the vector meson dominance
model, the most accurate value of the product $B(\phi\to e^+e^-)B(\phi\to
K_SK_L)=(9.85\pm0.03 \pm 0.10)\times10^{-5}$ has been obtained, which is
lower than the PDG value~\cite{pdg} by $1.6\sigma$.  The measured $\phi$
meson mass and width are consistent with the PDG values~\cite{pdg}. The best
description of the cross-section data near the $e^+e^-\to K_SK_L$ threshold
has been obtained with the model taking into account the final state
interaction. The significance of the FSI effect in our data is $3.2\sigma$.

\section{Acknowledgments}
This work was supported by the Russian Science Foundation Grant
No. 24-22-0203.

\end{document}